\documentclass[
 amsmath,amssymb,
 aps,
prb,
12pt]{revtex4-2}

\usepackage{float}

\usepackage{amsfonts, bbold}

\usepackage{hhline}
\usepackage{enumitem}
\usepackage{subcaption}
\usepackage{graphicx}
\usepackage{dcolumn}
\usepackage{slashed}
\usepackage{bm}
\usepackage{tabularx}
\usepackage{braket}
\usepackage{hyperref}
\usepackage{soul}
\usepackage{xcolor}

\begin{document}

\preprint{APS/123-QED}

\title{A Dark Matter Fermionic Quantum Fluid from Standard Model Dynamics}

\author{Stephon Alexander}
    \email{stephon\_alexander@brown.edu}
\author{Heliudson Bernardo}%
    \email{heliudson\_bernardo@brown.edu}
\author{Humberto Gilmer}
    \email{humberto\_gilmer@brown.edu}
\affiliation{%
 Department of Physics,\\
Brown University, Providence, RI 02912, USA
}%


\begin{abstract}

We present a model of dark matter as a superconducting fluid of Cooper pairs of right handed neutrinos. The superconducting dark matter is induced by attractive channels in the Standard Model Higgs interaction. We show that, for each case, the solution to the gap equation provides viable dark matter candidates for suitable chemical potential values. The mechanism yields an ultra-light BCS neutrino condensate with a mass of $m_{\rm DM} \sim 10^{-19} \text{eV}$. Both cosmological and particle physics constraints on the model lead to a connection between the number of effective relativistic species $N_{\rm eff}$, and the chemical potential and CMB temperature at the time of fermion creation. For certain values of the fermionic chemical potential in the early universe, we find a relation between the superconducting fermion and baryon densities, with implications for the coincidence between the dark matter and baryon densities in standard cosmology.  

\end{abstract}

\maketitle


\section{Introduction}
The existence of dark matter (DM) is now a settled component of the concordance model of cosmology and concrete evidence of physics beyond the Standard Model (SM) \cite{Bertone:2004pz}. Observations ranging from galactic rotation curves, large scale structure, to the cosmic microwave background (CMB) radiation have established evidence both for the existence of DM and its impact on visible matter, and on the cosmological history of the universe \cite{Arbey:2021gdg}. However, despite the robust evidence for its existence, the microphysical nature and identity DM are still unknown.

In this paper we propose a minimal model of dark matter which relies on the universality of the pairing mechanism of Bardeen-Cooper-Schrieffer (BCS) theory (Cooper pairs) induced by attractive interations given by either the Higgs or QCD sector of the Standard Model. The idea that the dark matter can be a BCS condensate was first discussed by the authors \cite{AC,Alexander:2018fjp,Garani}, and has been explored in other cosmological contexts \cite{Tong:2023krn}, however its connection to the Standard Model was unexplored.  In this manner, the dark matter is a stable massive Higgs mode of a superconducting state of fermions.  As we will see, the resulting dynamics predicts a new relationship between the concordance of the CMB temperature, $T_{\rm CMB}$ and the chemical potential of the fermions (at the time of their creation) which form the dark matter superconductor. 

Our description of the dark matter will have implications for extensions, if any, of the Standard Model and could give indications for further experimental detections to discriminate amongst various candidate models. Importantly, any viable microphysical description must satisfy a few criteria. (1) It must be dark, with a small a cross-section-to-mass ratio $\tfrac{\sigma}{M}\ll1$ for SM-DM processes. It is imperative that the microscopic model explain why ordinary SM interactions are absent or small. (2) The condensate must satisfy the observational constraints on dark matter, such as a small self-interaction rate and a dust-like equation of state.

Beyond these basic criteria, a successful microphysical description must further explain observational features. Both WMAP and more recently, Planck have measured a rough parity between the  dark matter energy density $\Omega_c$ and baryonic energy density $\Omega_b$, namely $\Omega_c\sim5\Omega_b$ \cite{WMAP:2006bqn,Planck:2018vyg}.  Classes of models exist which seek to explain this apparent coincidence. One such class is known as asymmetric dark matter (ADM) \cite{McDonald:2010toz, Davoudiasl:2012uw, Petraki:2013wwa}. These models postulate that the present-day DM density arises due to an asymmetry in the DM particle-antiparticle density, similar to the baryon asymmetry, and that these were generated by similar mechanisms. This microphysical model is contrasted with that of the WIMP dark matter \cite{Arcadi:2017kky}, in which the similarity in dark and visible matter densities must be taken as a coincidence \cite{McDonald:2010toz,McDonald:2011sv,Cui:2011ab}. Additionally, galactic DM distributions strongly suggest a DM candidate more complex than a collisionless WIMP. \cite{Salucci:2018hqu}

Certainly, ADM is not the only approach that invokes a common origin for baryonic and dark matter; instead, there exists the possibility that DM is not a separate fundamental field, but is instead a new, exotic state of visible, Standard Model matter. Such a proposal has precedent, with previous ideas ranging from quark nuggets, to strange matter (known as strangelets) \cite{Witten:1984rs, Farhi:1984qu} to QCD balls \cite{Zhitnitsky:2002qa}. Under the influence of an unspecified interaction, SM fermions may pair to form condensates that behave as scalar fields, analogously to the Cooper pairing \cite{PhysRev.104.1189} of BCS theory \cite{PhysRev.106.162, PhysRev.108.1175}. While there are proposals in which the condensate is accomplished via a dark gauge interaction \cite{Alexander:2018fjp, Alexander:2020wpm}, in this paper, the fermion interacts via SM boson mediation.  

The paper is organized as follows. In section \ref{sec:DM_superconductor}, we discuss the effective theory for the Higgs mode of the condensate and translate cosmological constraints on DM to constraints on its parameters. In section \ref{sec:neutrino_scenario}, we provide a specific implemention wherein the fermions giving rise to the condensate are BSM neutrinos, and in section \ref{sec:conclusion} we conclude with a discussion on future directions. Throughout the paper, we work with the mostly plus metric signature convention and assume natural units when otherwise not mentioned.

\section{Dark matter from superconductor states}\label{sec:DM_superconductor}


Here we discuss how DM can be identified as the Higgs mode of the condensate. We start from a generic effective theory for the interacting fermions and discuss the particular neutrino scenario in sections \ref{sec:neutrino_scenario}. The starting point is a Nambu-Jonas-Lasinio-like fermion Lagrangian \cite{Nambu:1961tp,Nambu:1961fr} with an attractive four-fermion interaction of the form which is induced from an attractive channel of fermions.
\begin{equation}\label{4fermionaction}
    \mathcal{L} = -\Bar{\psi} \left(\slashed\partial -m_{\psi} - \mu \gamma^0 \right)\psi + \frac{g^2}{M^2} \left(\Bar{\psi}\Gamma^A\psi\right)\left(\Bar{\psi}\Gamma_A\psi\right). 
\end{equation}
Introducing Hubbard-Stratonovich fields $\Phi_\alpha$ \cite{stratonovich1957method, Hubbard:1959ub}, we can write the theory above as one without a four-fermion interaction
\begin{align}
    \mathcal{L}_{\Phi} &= -\Bar{\psi} \left(\slashed\partial -m_{\psi} - \mu \gamma^0 \right)\psi\nonumber
    \\
    &+ \Phi_\alpha \left(\psi^\dagger \mathcal{O}^\alpha \psi^*\right) + \Phi_\alpha^* \left(\psi^\text{T}{\mathcal{O}^\alpha}^* \psi\right) - \frac{M^2}{g^2}\Phi_\alpha {\Phi^\alpha}^*.
\end{align}
where the $\mathcal{O}^{\alpha}$ is a generic spinor matrix structure, which we've left general for now. In subsequent sections, this will be specified. Since we wish to apply the following results for different attractive channels in different theories, we shall write the formal dependence on $\mathcal{O}_\alpha$ and the index $\alpha$ explicitly. The action corresponding to $\mathcal{L}$ can be recovered after using the equations of motion for $\Phi_\alpha$ and $\Phi^*_\alpha$.  

In the Nambu-Gorkov basis \cite{Gorkov1958, Nambu:1960tm}
\begin{equation}
    \Psi = \begin{pmatrix}
        \psi\\
        \psi^*
    \end{pmatrix},
\end{equation}
we have
\begin{equation}
    \mathcal{L}_{\Phi} = - \frac{1}{2}\Bar{\Psi} S^{-1}_F(\Phi_\alpha, m_{\psi}, \mu) \Psi - \frac{M^2}{g^2}\Phi_\alpha {\Phi^\alpha}^*,
\end{equation}
where 
\begin{align}
    &S_F^{-1}(\Phi_{\alpha}, m_{\psi}, \mu)\nonumber
    \\
    &= \begin{pmatrix}
        \gamma^\mu \partial_\mu - m_{\psi} - \mu \gamma^0 & \Phi_{\alpha} i\gamma^0 \mathcal{O}^\alpha \\
        \Phi_{\alpha}^* i\gamma^0 {\mathcal{O}^\alpha}^* & i\gamma^0 {\gamma^\mu}^\text{T}i\gamma^0 \partial_\mu - m_{\psi} - \mu \gamma^0
    \end{pmatrix}
\end{align}
In the mean-field approximation, an effective action for $\Phi$ is obtained after integrating out the fermionic fields,
\begin{align}\label{appendix:Seff}
    &e^{i S_{\text{eff}}(\Phi_{\alpha}, \Phi_{\alpha}^*)}\nonumber
    \\
    &= \int \left[\mathrm{d}\Bar{\Psi}\mathrm{d}\Psi\right]e^{-i \int \mathrm{d}^4 x\left(\frac{1}{2}\Bar{\Psi} S^{-1}_F(\Phi_{\alpha}, m_{\psi}, \mu) \Psi + \frac{M^2}{g^2}\Phi_{\alpha} {\Phi^{\alpha}}^*\right)}
\end{align}
and the gap equation is 
\begin{equation}\label{appendix:gapeq}
    \frac{\delta S_{\text{eff}}}{\delta \Phi_{\alpha}} =0,
\end{equation}
and its solution fixes the value of $\Phi_{\alpha} = \Bar{\Phi}_\alpha(\mu, m_{\psi}, M^2/g^2)$ in terms of the parameters of the original Lagrangian. The gap $\Bar{\Phi}_\alpha$ determines the value of the condensate:
\begin{equation}\label{fermionicvev}
    \langle \psi^\text{T} \mathcal{O}^\alpha \psi\rangle \propto \Bar{\Phi}^\alpha.
\end{equation}
One core fact of BCS theory is that there is always a solution to the gap equation whenever the interaction term corresponds to an attractive channel \cite{Joe, Shankar:1993pf}. Moreover, any repulsive channel becomes irrelevant to the computation of the gap. This can be more easily seen from a Wilsonian perspective, in which modes away from the Fermi surface are integrated out, and repulsive interaction terms become irrelevant at those energies \cite{Joe, Shankar:1993pf}. 

In this paper, we propose to identify the massive, radial mode $\rho_\alpha(x^\mu)$ of the gap as the dark matter candidate. To get the microscopic DM parameters, such as mass and self-interaction couplings, we will then need an effective theory for fluctuations around $\Bar{\Phi}_\alpha$:
\begin{equation}
    \Phi_\alpha(x^\mu) = \left[\Bar{\Phi}_\alpha + \rho_\alpha(x^\mu)\right] e^{i\phi(x^\mu)},
\end{equation}
where $\phi$ is the Goldstone mode of the condensate. This mode will be associated to a symmetry breaking by \eqref{fermionicvev}. The effective field theory for the fluctuation modes was studied in \cite{Anglani:2011cw} by perturbing around the ground state of the resulting effective action in \eqref{appendix:Seff}. In terms of the canonically normalized field $\varphi = \left(\sqrt{3}/2\pi\right)\left(\mu/\Phi\right)\rho$, the Lagrangian for the Higgs mode is then \cite{Anglani:2011cw, Alexander:2018fjp}
\begin{equation}
    \mathcal{L} = -\frac{1}{2}(\partial\varphi)^2 - 16 \bar{\Phi}^2 \varphi^2 - \sum_{n\geq 2} \tilde{c}_n \frac{\bar{\Phi}^2}{\mu^{n-2}}\varphi^n,
\end{equation}
where $\Tilde{c}_n$ are order unity constants. From the Lagrangian above one identifies the DM mass $m$ and quartic coupling $\lambda$ as
\begin{equation}\label{DM_mass_int}
    m = 4\sqrt{2}\Bar{\Phi}, \quad \lambda \simeq \frac{\Bar{\Phi}^2}{\mu^2}.
\end{equation}

In the scenario explored in section \ref{sec:neutrino_scenario}, we introduce RH neutrions on top of the SM leptons and quarks (however, the interaction giving rise to the four-fermion term in \eqref{4fermionaction} still comes from the SM bosons). We shall assume these neutrinos were produced in the early universe, possibly during leptogenesis, or during reheating \cite{Adshead:2015jza}. For generality, we forgo the details of production but rather comment on constraints on the temperature and chemical potential of the fermionic distribution. To get a Fermi surface, the fermions should be in the degenerate limit, with chemical potential much larger than the temperature, $\mu/T \gg 1$. Motivated by the Sakharov conditions \cite{Sakharov:1967dj} and electroweak baryogenesis \cite{Cohen:1993nk,Riotto:1998bt,Trodden:1998ym,Morrissey:2012db}, these fermions may have been produced out of equilibrium and we can neglect the temperature corrections at leading order.

\section{Right-handed neutrino condensate}\label{sec:neutrino_scenario}

We propose a concrete scenario that realizes dark matter from fermionic condensates, namely a right-handed sterile neutrino\footnote{By ``sterile'' we mean that the right-handed neutrinos are singlets of the SM gauge groups. Strictly speaking, in the neutrino case we have a superfluid condensate, but our formalism generally accommodates superconductor states.}. This is a natural extension of the SM fermion sector, with no changes in its bosonic sector. The notion of a neutrino condensate mediated by a scalar field between left- and right-handed neutrinos is not a new idea. However, most treatments of this idea miss an important subtlety which opens the possible parameter space. Yukawa interactions generically couple left- and right-handed fermion fields; this would in turn lead to the idea that any resulting condensate should consist of left-right (LR) bound states. However, it has been shown that the Yukawa interaction can mediate the formation of a condensate in both the $t$ channel \cite{Kapusta:2004gi} (the expected LR condensate) and the $s$ channel  \cite{Pisarski:1999av, Chodos:2020dgi} (leading to the possible formation of LL and RR condensates). 

\begin{figure}[h]
\centering
\includegraphics[width = 0.3\textwidth]{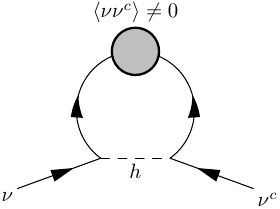}
\caption{\label{fig:4Fermi_gap} Diagrammatic representation of a LL or RR gap arising from the Yukawa interaction between left- and right-handed neutrinos.  The Higgs (dashed line) induces an attractive four-fermi interaction. The grey blob denotes a sum over neutrino ladder diagrams.}
\end{figure}

Given that the replusive $Z$ exchange is irrevelant at finite density, the relevant Lagrangian for the neutrino-Higgs sector, written in terms of Weyl spinors, is
\begin{equation}
    \mathcal{L} = -\nu^{\dagger}\bar{\sigma}\cdot\partial\nu  - \bar{\nu}\sigma\cdot\partial\bar{\nu}^{\dagger} + y_{\nu} L\cdot H\bar{\nu} + \text{H.c.}
\end{equation}
where $\nu$ is the left-handed (Standard Model) neutrino and $\bar{\nu}$ is the right-handed neutrino. Upon inserting the Higgs vev and integrating out the resulting interaction, we obtain the effective Lagrangian for Dirac neutrinos at finite densities and below the Higgs mass scale,
\begin{equation}
    \mathcal{L} = -\Bar{\psi} \left(\slashed\partial -m - \mu \gamma^0 \right)\psi + \frac{y_\nu^2}{M_h^2} \left(\Bar{\psi}\psi\right)\left(\Bar{\psi}\psi\right),
\end{equation}
where  $\psi^{\rm T} = (\nu, \bar{\nu}^{\dagger})$ is the Dirac spinor containing the neutrino fields, $y_\nu$ is the neutrino Yukawa coupling and $M_h$ is the Higgs mass. We will see that condensates containing both the same and different chiralities are possible using this effective Lagrangian. The resulting gap may be represented diagrammatically as in Figure \ref{fig:4Fermi_gap}.
This action is equivalent to \eqref{4fermionaction} with one quartic-coupling operator $\Gamma^1 = \mathbb{1}$ and $g^2/M^2 = y_\nu^2/M_h^2$. In the appendix \ref{appendix_A}, we use a Fierz rearrangement to show that the only operators $\mathcal{O}_\alpha$ corresponding to attractive channels are $\gamma^2 \gamma_5/4$ and $\gamma^1 \gamma^3/4$, such that there are two possible gaps:
\begin{equation}
    \langle \psi^\dagger \gamma^2 \gamma_5 \psi \rangle \propto \Phi_{\rm LR}, \quad \langle \psi^\dagger \gamma^1 \gamma^3 \psi\rangle \propto \Phi_{\rm LL + RR} .
\end{equation}
The former corresponds to an LR condensate while the latter to an $\text{LL}+\text{RR}$ condensate. Since the SM left-handed neutrinos were in thermal equilibrium at early times, the $\text{LL}$ part does not contribute to the gap, such that $\langle \psi^\dagger \gamma^1 \gamma^3 \psi \rangle \to \langle \bar{\nu} i\sigma^2\bar{\nu}^{\dagger} \rangle \propto \Phi_{\rm RR}$. Moreover, in \cite{Chodos:2020dgi}, it is shown that the condensate made of neutrinos of the same chirality is energetically favorable over $\Phi_{\rm LR}$. The solution to the gap equation \eqref{appendix:gapeq} is \cite{Chodos:2020dgi}
\begin{equation}
    \Phi_{\rm RR} = 2 \sqrt{M_h(\mu - m_\psi)}\;e^{-\frac{\pi^2}{2\kappa_R^2 \mu\sqrt{\mu^2-m_\psi^2}}},
\end{equation}
where $\kappa_R^2 \simeq y_\nu^2/M^2_h$. The exponential dependence on the inverse coupling squared $\Phi \propto e^{-1/y_\nu^2}$ is expected from BCS theory. Note that the condensate breaks the lepton number symmetry spontaneously. 

In the following section, we will find a wide range of parameters compatible with the above constraints, generating a broad spectrum of DM masses. In general, we expect phenomenology akin to the superfluid DM (see e.g. \cite{Arguelles:2023nlh} and references therein); namely, we expect the large-scale evolution of the DM background energy density to be described by the fermionic one, while the scalar nature of the Higgs modes will play a leading role on galactic and sub-galactic scales, evading constraints on fermionic DM candidates, such as the Tremaine-Gunn bound \cite{PhysRevLett.42.407}.

\section{Cosmological constraints}
\label{sec:neutrino_scenario}

The energy density of the fermion fluid will be constrained by cosmological history. In the standard cosmic history context, adding an extra fluid before BBN corresponds to a change in the effective number of relativistic species, $N_{\text{eff}}$, which is set by the relativistic particle content of the SM. A shift $\Delta N_{\text{eff}}$ of this value is constrained by cosmological observations \cite{Planck:2018vyg,Workman:2022ynf}:
\begin{equation}\label{appendix:deltaNeff}
    \Delta N_{\text{eff}} < 0.17 \quad (\text{CMB} + \text{BAO})\; \text{at}\; 95\% \;\text{CL}.
\end{equation}

We now use this constraint to find a maximum value for the Fermi energy of the degenerate gas. In the limit $\mu/T \gg1$, the number and energy densities for a Fermi (ideal) gas are
\begin{align}
    n_\psi &= \frac{g}{6\pi^2}\left(\mu^2 - m_\psi^2 \right)^{3/2},
    \\ \label{appendix:densities}
    \rho_\psi &= \frac{g}{16\pi^2}\left[\mu\left(2\mu^2 - m_\psi^2\right)\sqrt{\mu^2 -m_\psi^2}- m_\psi^4 \text{arcosh}\left(\frac{\mu}{m_\psi}\right)\right],
\end{align}
where $g$ is the number of degrees of freedom of each fermionic particle; each extra fermion flavor contributes a value of 2 to $g$. Note that for $\mu \gg m_\psi$, we have $n_\psi \propto \mu^3$ and $\rho_\psi \propto \mu^4$.

In an Friedmann-Lema\^itre-Robertson-Walker (FLRW) background, the chemical potential dependence of the scale factor can be obtained from particle number conservation. Using the conservation of the cosmic fluid's entropy to write scale factor dependence in terms of temperature, we have (see \cite{Trautner:2016ias} for details)
\begin{equation}\label{appendix:mu_scaling}
    \mu(T) = m_\psi\sqrt{1+\left(\frac{g_{*S}(T)}{g_{*S}(T_\psi)}\right)^{2/3}\left(\frac{\mu(T_\psi)^2 - m_\psi^2}{T_\psi^2}\right)\left(\frac{T}{m_\psi}\right)^2 },
\end{equation}
where $g_{*S}(T)$ is the effective number of degrees of freedom in entropy at temperature $T$ and $T_\psi$ is the temperature of the cosmic fluid when the non-thermal fermions were created. The evolution of the fermionic energy density is given by $\rho_{\psi}(T) = \rho_\psi(\mu(T))$. At early times, when $T\gg m_\psi$, $\rho_\psi$ redshifts as radiation,
\begin{equation}
\label{rho_scaling}
    \rho_{\psi}(T) \simeq \frac{g}{8\pi^2}\left(\frac{\mu(T_\psi)}{T_\psi}\right)^4 \left(\frac{g_{*S}(T)}{g_{*S}(T_\psi)}\right)^{4/3} T^4,
\end{equation}
where we assumed $\mu(T_\psi) \gg m_\psi$. After the temperature of the cosmic radiation drops below $m_\psi$, $\rho_\psi \propto T^3$ and we have a dust-like equation of state.

The shift in $N_{\text{eff}}$ during BBN is then \cite{Chen:2015dka}
\begin{equation}
    \Delta N_{\text{eff}} = \frac{8}{7}\frac{30 g}{16\pi^4}\left(\frac{\mu(T_\psi)}{T_\psi}\right)^4 \left(\frac{g_{*S}(T_{\text{BBN}})}{g_{*S}(T_\psi)}\right)^{4/3}.
\end{equation}
As long as $\Delta N_{\text{eff}}$ is less than the upper bound in \eqref{appendix:deltaNeff}, the mild cosmic history modification due to the presence of the fermion fluid satisfies the current cosmological constraints. Hence, \eqref{appendix:deltaNeff} implies that $\mu(T_\psi)$ should be at most of order $T_\psi$:
\begin{equation}\label{appendix:muconstraint}
    \frac{\mu(T_\psi)}{T_\psi} \lesssim \frac{1.22}{g^{1/4}}.
\end{equation}

The DM self-interaction cross-section $\sigma$ is constrained to satisfy $\sigma/m < 0.1 \;\text{cm}^2/\text{g}$ (see \cite{Tulin:2017ara} and references therein). Using the expression for the mass and quartic coupling \eqref{DM_mass_int}, we have
\begin{equation}
    \frac{m}{\text{eV}} \lesssim 10^{-19} \left(\frac{\mu_0}{\text{eV}}\right)^4
\end{equation}
where $\mu_0$ represents the Fermi energy today. There is a range of possible DM masses depending on the chemical potential right after the fermion production. From \eqref{appendix:mu_scaling}, we have
\begin{align}\label{appendix:DMmassbound1}
    \frac{m}{\text{eV}} &\lesssim 10^{-19}\left(\frac{m_\psi}{\text{eV}}\right)^4\times\nonumber
    \\
    &\left[1+\left(\frac{g_{*S}(T_0)}{g_{*S}(T_\psi)}\right)^{2/3}\left(\frac{\mu(T_\psi)^2 - m_\psi^2}{T_\psi^2}\right)\left(\frac{T_0}{m_\psi}\right)^2\right]^2
\end{align}
Using the constraint \eqref{appendix:muconstraint} with $\mu(T_\psi) \gg m_\psi$, we get
\begin{equation}
    \frac{m}{\text{eV}} \lesssim 10^{-19}\left(\frac{m_\psi}{\text{eV}}\right)^4\left[1+\frac{1.22}{g^{1/4}}\left(\frac{g_{*S}(T_0)}{g_{*S}(T_\psi)}\right)^{2/3}\left(\frac{T_0}{m_\psi}\right)^2\right]^2,
\end{equation}
which for $T_0 \ll m_\psi $ gives
\begin{equation}
\label{DMmassbound2}
    \frac{m}{\text{eV}} \lesssim 10^{-19} \left(\frac{m_\psi}{\text{eV}}\right)^4.
\end{equation}
Hence, for a wide range of fermion masses, this model satisfies the self-interaction constraints provided DM is ultralight \cite{Hu:2000ke, Hui:2016ltb} and consequently the DM phenomenology resembles the one for ultralight axions (see \cite{Marsh:2015xka,Ferreira:2020fam, Hui:2021tkt} and references therein).
However, from the specific form of the gap, the DM mass can take on a wider range of values, corresponding to anything from ultralight DM to WIMPs \cite{Roszkowski:2017nbc} (and references therein), all depending primarily on the chemical potential. The ultralight DM region is phenomenologically interesting because it would reproduce many aspects of superfluid dark matter (see e.g. \cite{Khoury:2021tvy} and references therein).


We can estimate which values of $\mu_\psi$ can explain the coincidence $\Omega_c \sim 5 \Omega_b$ after assuming that the number density of fermions right after they are created $n_\psi(T_\psi)$ is of the same order of the net baryon number density $n_{b} = \eta n_\gamma$ at that time, where $\eta$ is the baryon-to-photon ratio and $n_\gamma$ is the photon number density. The present baryonic density parameter is related to $\eta(T_0)$ as
\begin{equation}
    \Omega_b h^2 \simeq 3.66 \times 10^7 \eta,
\end{equation}
and BBN an CMB data implies $\eta(T_0) \simeq 6 \times 10^{-10}$ \cite{Planck:2018vyg,Workman:2022ynf, Yeh:2022heq}. So, the baryonic density can be explained as long as we reproduce the observed value for $\eta(T_0)$. From
\begin{equation}
    n_b(T_\psi) = \xi_\psi(T_\psi) n_\psi (T_\psi),
\end{equation}
and the expression for the fermionic number density in \eqref{appendix:densities}, we get
\begin{equation}
    \left(\frac{\mu_\psi^2 - m_\psi^2}{T^2_\psi}\right)^{3/2} =\frac{12\zeta(3)}{g}\frac{\eta(T_\psi)}{\xi_\psi}.
\end{equation}
From
\begin{equation}
    \eta = \frac{n_\text{B}}{n_\gamma} = \frac{\pi^4}{45 \zeta(3)}g_{*S}(T)\frac{n_\text{B}}{s},
\end{equation}
where $s$ is the radiation entropy density, we have 
\begin{equation}
    \left(\frac{\mu_\psi^2 - m_\psi^2}{T^2_\psi}\right)^{3/2} =\frac{12\zeta(3)}{g}\frac{g_{*S}(T_\psi)}{g_{*S}(T_0)}\frac{\eta(T_0)}{\xi_\psi}.
\end{equation}
For large $\mu_\psi$ and $g_{*S}(T_\psi) \sim \mathcal{O}(10^2)$, 
\begin{equation}\label{appendix:coincidence_constraint}
    \frac{\mu_\psi}{T_\psi} \simeq \frac{1}{(g\xi_\psi)^{1/3}} 6 \times 10^{-3}.
\end{equation}
As $\xi_\psi,\, g \sim \mathcal{O}(1)$ we conclude that the relation between the baryonic and cold matter densities is reproduced provided $\mu_\psi$ is around 3 orders of magnitude smaller than the cosmic radiation temperature when the fermions are created. Note that \eqref{appendix:coincidence_constraint} and \eqref{appendix:muconstraint} are compatible; hence, under the assumptions of this section, a degenerate gas of fermions in the early universe can indeed explain the $\Omega_c \sim 5 \Omega_b$ coincidence today.

For tuned values of the chemical potential at early times, the condensate's Higgs mode reproduces the dark nature of DM. For $m_\psi \approx 0.1 \, \text{eV}$ and $\mu_\psi \approx 1.5 \times 10^{14}\; \text{GeV}$, we have $\Phi_{\rm RR} \approx 10^{-20} \; \text{eV}$, such that the bound \eqref{appendix:DMmassbound1} is compatible with \eqref{DM_mass_int}. At later times, the cosmological bounds will continue to be satisfied, because as $\mu$ gets closer to $m_\psi$, $\Phi_{\rm RR}$ is further suppressed. Note that the gap depends mainly on the chemical potential value.
The matter of generating such a chemical potential is an open question in this work, but possible means of addressing this question have been examined in e.g. \cite{Adshead:2015jza}, wherein the inflaton's VEV can act as a fermionic chemical potential.

\section{Discussion and Conclusion}\label{sec:conclusion}
In this work we have presented an analysis regarding whether dark matter states are viable within the structure of the Standard Model. Although it is not possible, due to precision data, to construct dark states using only Standard Model fields, it is possible to do so within the gauge structure of the Standard Model, by augmenting the matter sector with well-motivated additional fermionic fields, e.g. the sterile neutrino in this work. Other scenarios would have their own unique aspects, but much of the physics common to this and other scenarios arises from the well-known physics of superconductivity.

Postulating a degenerate fermionic background, for sufficient values of chemical potential, the formation of a condensate is inevitable given an attractive interaction. This presents the possibility that what we have heretofore called dark matter is simply a particular state of the fermionic species, mediated by the known interactions of the Standard Model. In this case, the dark matter particle would be the Higgs mode of the resulting superconducting state, similar to how the Higgs boson detected at the LHC is but the excitation of the background Higgs field. 

The necessary chemical potential in the neutrino case is adjusted to account for the fact that the neutrino mass is bounded from above and that the interaction is mediated by the Higgs which is relatively heavy for such an interaction. Due to the constraints on the neutrino mass and the Higgs mass, the possible values for the DM particle are fixed primarily by the chemical potential and it is thus possible in this case to generate DM masses in the ultralight range.

In our proposal, there is only one fluid, fundamentally made of fermions, but in a state that gives rise to different modes (Goldstone and Higgs excitations). The Goldstone mode is massless and so it redshifts as radiation right after the formation of the condensate, hence it is not relevant for the background energy density (similar to the model discussed in \cite{Alexander:2018fjp}). The DM particle consists of Higgs mode excitations of the background superconducting state. Since we have local energy conservation, the DM density can be obtained by redshifting the fermionic energy density \eqref{appendix:densities} in the fermions right after their creation. Alternatively, we can calculate $\Omega_c$ at the time of condensate formation and compare it with Eq. \eqref{appendix:densities}. In that sense any bounds on $\Omega_{c}$ will translate into bounds on $\rho_\psi$. For instance, to get Eq. \eqref{appendix:coincidence_constraint} we used the observed value of $\Omega_b$, its relation with $\Omega_{c}$, and the value of the baryon-to-photon ratio at early times.

It is worth commenting that the matter-radiation equality scale $T_{\rm eq}\sim 0.3 \; \text{eV} $ is of the order of the cosmological upper bound on the sum of neutrino masses, $\sum_i m_{\nu, i} < 0.13 \; \text{eV}$ \cite{Planck:2018vyg,Workman:2022ynf}. For the neutrino case, $\rho_\psi$ starts to redshift as dust when $T \sim m_\nu $, i.e. just before $z_{\rm eq}$. Or, equivalently, neutrinos having an $\mathcal{O}(10^{-1}) \; \text{eV}$ mass may be the reason why $z_{\rm eq} \sim \mathcal{O}(10^3)$. Another possible line of investigation is whether the condensate can play any role in solutions to the Hubble tension (see \cite{DiValentino:2021izs, Hu:2023jqc} for reviews), given the characteristic scale (eV) that new physics should have in order to solve the tension \cite{Kamionkowski:2022pkx}.

An important aspect left for future work is the phenomenology of this model, specifically examining such aspects as the core-cusp problem or the baryonic Tully-Fisher relation. For a wide range of parameters, we expect the small-scale phenomenology of the condensate to be similar to the superfluid DM case, thus avoiding usual astrophysical constraints on fermionic particle dark matter candidates \cite{Arguelles:2023nlh}. Other phenomena associated with superfluid DM, such as vortices \cite{Alexander:2019qsh, Alexander:2019puy, Alexander:2021zhx}, are also possible and worth exploring in this context. It will be important to distinguish this model from a fundamental ultralight scalar dark matter candidate.  One possibility for future work is to study the possible neutrino oscillations induced by the background neutrino condensate as neutrinos propagate through it. Another possibility for implementing the condensate from attractive channels within the SM arises for the case of the color-flavor locking phase of VLQs due to single-gluon exchange. We leave the exploration of this possible scenario to a future work. From a more theoretical perspective, one could also investigate how the four-fermion interaction affects the fermionic densities and get a fully-fledged expression for the condensate contribution to the dark matter density. Another direction would be explicitly constructing the baryogenesis model that gives origin to a degenerate fermion background with suitable chemical potential.

\begin{acknowledgments}

We thank Chris Cappiello, Ian Dell'Antonio, Loukas Gouskas, Leah Jenks, Greg Landsberg, Evan McDonough, Philip Phillips, Michael Peskin, and David Spergel for useful discussions. We thank Jorge Sofo for inspiring SA for pursuing this project. HBG is supported by the National Science Foundation, MPS Ascending Fellowship, Award 2213126.

\end{acknowledgments}

\appendix

\section{Fierz rearrangement}\label{appendix_A}

We want to express the bilinear squared $(\Bar{\psi}\psi)^2$ as a product of $\psi^\dagger \psi^\dagger$ and $\psi \psi$ to make particle-particle pairing in the mean field approximation. We have
\begin{equation}\label{Fierzrearrange}
    (\Bar{\psi} \psi)^2 = (\psi^\dagger i \gamma^0 \psi)(\psi^\dagger i \gamma^0 \psi) =\sum_{A,B}^{16}\xi_{AB}(\psi^\dagger \Theta^A \psi^*)(\psi^\text{T} \Bar{\Theta}^B \psi), 
\end{equation}
where $\Theta^A$ are the elements of the Clifford algebra basis and only the six antisymmetric among those contribute to the summation. Typically, one works on the basis
\begin{align}
    \Gamma^A &= \{\mathbb{1}, \gamma^\mu, \gamma^{\mu\nu}, \gamma^{\mu\nu\rho}, \gamma^{\mu\nu\rho\sigma}\},\nonumber\\ \Gamma_A &= \{\mathbb{1}, \gamma_\mu, \gamma_{\nu\mu}, \gamma_{\rho\nu\mu}, \gamma_{\sigma\rho\nu\mu}\}, \quad \sigma > \rho > \nu> \mu,
\end{align}
where $\gamma^{\mu_1 \cdots \mu_r} = \gamma^{[\mu_1}\dots\gamma^{\mu_r]}$. These matrices have normalization such that
\begin{equation}\label{Gammanormalization}
    \text{tr}\left(\Gamma^A \Gamma_B\right) = 4\delta^A_B.
\end{equation}
Note that $\{\Gamma^A\}$ is the basis for the vector times dual-vector representation. This can be seen from the fact that $\bar{\psi}\Gamma^A \chi$ is a scalar. On the other hand, $\Theta^A$ ($\Bar{\Theta}^B$) denotes the basis elements in the vector-vector (dual-vector times dual-vector) representation, such that $\Bar{\chi} \Theta^A \Bar{\psi}^\text{T} = (\psi^\text{T}\Bar{\Theta}^A\chi)^\dagger$ is a scalar. This relation also implies that $\Bar{\Theta}^A =(i\gamma^0)^*{\Theta^A}^\dagger (i\gamma^0)^*$. The basis in both representations are related by $\Theta^A = \Gamma^A C$, where $C$ is the unitary charge conjugation matrix. Such a matrix enters in the symmetry property of $\Gamma^A$:
\begin{equation}
    {\Gamma^A}^\text{T} = t_0 t_r C \Gamma^A C^{-1},
\end{equation}
where $t_r = \pm1$ depending on the rank of $\Gamma^A$. Choosing $t_0 = 1 = -t_1$, we have $t_2 =-1$ and $t_3= 1= t_4$. In a representation in which ${\gamma^\mu}^\dagger = \gamma^0\gamma^\mu \gamma^0$ (hermitian representation) we have
\begin{equation}
    {\gamma^\mu}^* = -t_0t_1 B \gamma^\mu B^{-1}, \quad B = i t_0 C\gamma^0.
\end{equation}
Using this result, it is straightforward to show that
\begin{equation}
    {\Gamma^A}^\dagger = t_1 t_r \gamma^0 \Gamma^A \gamma^0,
\end{equation}
and hence 
\begin{equation}
    \Bar{\Theta}^A = -t_0t_{r(A)}C^\dagger \Gamma^A.
\end{equation}
One can show that, with the choice of basis above, we have the normalization
\begin{equation}
    \text{tr}\left(\Theta^A \Bar{\Theta}_B\right) = -t_0 t_{r(B)} \text{tr}\left(\Gamma^A \Gamma_B\right) = -t_0 t_{r(B)}4 \delta^{A}_B.
\end{equation}

To get the typical particle-antiparticle Fierz rearrangement, we start with the identity
\begin{equation}
    \left(\Gamma^C\right)_{ij} \left(\Gamma^D\right)_{k\ell} = \sum_{A,B} C^{CD}_{AB} \left(\Gamma^A\right)_{i\ell} \left(\Gamma^B\right)_{kj}
\end{equation}
and multiply it by $\left(\Gamma_E\right)_{jk} \left(\Gamma_F\right)_{li}$ to find
\begin{equation}\label{Ctraces}
    C^{CD}_{EF} = \frac{1}{16}\text{tr}\left(\Gamma^C \Gamma_E \Gamma^D \Gamma_F\right).
\end{equation}
Similarly, to get the particle-particle rearrangement, we start from
\begin{equation}\label{particleparticlerearrange}
        \left(\Gamma^A\right)_{ij} \left(\Gamma^B\right)_{kl} = \sum_{C,D} D^{AB}_{CD} \left(\Theta^C\right)_{il} \left(\Bar{\Theta}^D\right)_{kj}.
\end{equation}
To find
\begin{equation}
    D^{AB}_{FE} = \frac{1}{16} \text{tr}\left(\Gamma^A \Theta_E {\Gamma^B}^\text{T} {\Bar{\Theta}}_F\right).
\end{equation}
So, from the relation between $\Theta$ and $\Gamma$, we have
\begin{align}
    D^{AB}_{FE} &= -\frac{1}{16} t_{r(B)}t_{r(F)}\text{tr}\left(\Gamma^A \Gamma_E\Gamma^B \Gamma_F\right)\nonumber
    \\
    &= -t_{r(B)}t_{r(F)} C^{AB}_{EF}.
\end{align}

A glance at \eqref{Fierzrearrange} reveals that $\xi_{AB} = -D^{00}_{AB} = -t_{r(F)}C^{00}_{BA}$. We now compute $C^{00}_{BA}$. Firstly, note that the trace of an odd number of $\gamma^\mu$ matrix vanishes and so will the trace of $\Gamma^A$'s with unmatched rank. Thus $C^{00}_{BA} \propto \delta_{BA}$. To compute the traces in \eqref{Ctraces}, it is convenient to write the basis as $\Gamma^A = \{\mathbb{1}, \gamma^\mu, \gamma^{0i}, \gamma^{ij}, \gamma^\mu \gamma_5, \gamma_5\}$ with $i<j$, where $\gamma_5 = -i\gamma^{0}\gamma^1 \gamma^2 \gamma^3$. Note that for $\Gamma_A$, the rank-2 elements have opposite index ordering compared to rank-2 $\Gamma^A$ elements, otherwise the elements will not be normalized as in \eqref{Gammanormalization}. For instance,
\begin{equation}
    \left.\text{tr}(\gamma^{ij}\gamma_{k\ell})\right|_{\substack{j>i \\ k>\ell}} = 4\left(\eta^{ij}\eta_{k\ell}- \delta^i_k \delta^j_{\ell} + \delta^i_{\ell} \delta^j_k\right)_{\substack{j>i \\ k>\ell}} = 4 \delta^i_{\ell} \delta^j_k
\end{equation}
would have an opposite sign if $\ell>k$. With that in mind, we collect the results for $D^{00}_{AB}$:
\begin{align}
    D^{00}_{\mathbb{1}\mathbb{1}} &= t_0C^{00}_{\mathbb{1}\mathbb{1}}= -\frac{1}{4},\\
    D^{00}_{00} &= t_1 C^{00}_{00} = -\frac{1}{4}, \\
    D^{00}_{ij} &= t_1 C^{00}_{ij} = - \frac{1}{4}\eta_{ij},\\
    D^{00}_{i0\;j0} &= t_2 C^{00}_{i0\;j0} = -\frac{1}{4}\eta_{ij}, \\
    D^{00}_{ij\; k\ell} & = t_2 C^{00}_{ij\; kl}=  -\frac{1}{4}\eta_{ik}\eta_{j\ell}, \\
    D^{00}_{05\;05} &= t_3 C^{00}_{05\;05}=\frac{1}{4}, \\
    D^{00}_{i5\;j5} &= t_3 C^{00}_{i5\;j5}=\frac{1}{4}\eta_{ij},\\
    D^{00}_{55} &= t_4 C^{00}_{55}=\frac{1}{4}.
\end{align}

Using these results into \eqref{particleparticlerearrange}, we find
\begin{align}
    \left(\gamma^0\right)_{ij}\left(\gamma^0\right)_{k\ell} &= -\frac{1}{4}\left(C\right)_{ik} \left(-C^\dagger\right)_{\ell j} - \frac{1}{4}\left(\gamma^0 C\right)_{ik} \left(C^\dagger\gamma^0\right)_{\ell j}\nonumber\\
    &- \frac{1}{4}\eta_{mn}\left(\gamma^m C\right)_{ik} \left(C^\dagger \gamma^n\right)_{\ell j} \nonumber\\
    &-\frac{1}{4} \eta_{mm}\left(\gamma^{0m}C\right)_{ik} \left(C^\dagger\gamma^{0n}\right)_{\ell j} \nonumber\\
    &-\frac{1}{4}\eta_{mn}\eta_{pq}\left(\gamma^{mq}C\right)_{ik} \left(C^\dagger \gamma^{np}\right)_{\ell j}+\nonumber\\
    &+\frac{1}{4}\left(\gamma^0 \gamma_5 C\right)_{ik} \left(-C^\dagger \gamma^0 \gamma_5\right)_{\ell j}\nonumber\\
    &+ \frac{1}{4}\eta_{mn}\left(\gamma^{m}\gamma_5 C\right)_{ik} \left(-C^\dagger \gamma^n \gamma_5\right)_{\ell j}+\nonumber\\
    &+ \frac{1}{4}\left(\gamma_5 C\right)_{ik} \left(-C^\dagger \gamma_5\right)_{\ell j}.
\end{align}
When we multiply this expansion by $\psi^\dagger_i \psi_k^\dagger \psi_{\ell} \psi_j$ to get \eqref{Fierzrearrange}, only the terms with antisymmetric $\Theta^A$ will contribute. The symmetry property of $\Theta^A$ depends on its rank:
\begin{equation}
    {\Theta^A}^\text{T} = \left(\Gamma^A C\right)^\text{T} = t_r C^2 \Gamma^A C = t_r C^2 \Theta^A.
\end{equation}
We shall work concretely in the Weyl representation, where $C = -i\gamma^0 \gamma^2$. In that case, $C^2 = -\mathbb{1}$ and so $\Theta^A$ is antisymmetric for ranks such that $t_r = 1$ which correspond to $r = \{0, 3, 4\}$. Then we are interested in four terms in the expansion above:
\begin{align}
    \left(\gamma^0\right)_{ij}\left(\gamma^0\right)_{k\ell} &= -\frac{1}{4}\left(C\right)_{ik} \left(-C^\dagger\right)_{\ell j} \nonumber\\
    &- \frac{1}{4}\left(\gamma^0 \gamma_5 C\right)_{ik} \left(-C^\dagger \gamma^0 \gamma_5\right)_{\ell j} +\nonumber\\
    &+\frac{1}{4}\eta_{mn}\left(\gamma^{m}\gamma_5 C\right)_{ik} \left(-C^\dagger \gamma^n \gamma_5\right)_{\ell j}\nonumber\\
    &+ \frac{1}{4}\left(\gamma_5 C\right)_{ik} \left(-C^\dagger \gamma_5\right)_{\ell j} + \cdots \;.
\end{align}
Explicitly in terms of $\gamma$ matrices, we have
\begin{align}
    \left(i\gamma^0\right)_{ij}\left(i\gamma^0\right)_{k\ell} &= -\frac{1}{4}\left(\gamma^0\gamma^2\right)_{ik} \left(\gamma^0 \gamma^2\right)_{\ell j}\nonumber\\
    &- \frac{1}{4}\left(\gamma^2 \gamma_5 C\right)_{ik} \left(\gamma^2 \gamma_5\right)_{\ell j} -\nonumber\\
    &-\frac{1}{4}\left(\gamma^{3}\right)_{ik} \left(\gamma^3\right)_{\ell j}- \frac{1}{4}\left(\gamma^0\gamma_5 \right)_{ik} \left(\gamma^0 \gamma_5\right)_{\ell j} \nonumber\\
    &-\frac{1}{4}\left(\gamma^{1}\right)_{ik} \left(\gamma^1\right)_{\ell j} -\nonumber\\
    &-\frac{1}{4}\left(\gamma^1\gamma^{3}\right)_{ik} \left(\gamma^1\gamma^3\right)_{\ell j} +\cdots \;.
\end{align}
Finally, we express the particle-particle Fierz rearrangement in a manifestly real way as
\begin{align}
    (\Bar{\psi}\psi)^2 &= \frac{1}{4}\left(\psi^\dagger \gamma^0\gamma^2 \psi^*\right) \left[\psi^\text{T} \left(\gamma^0\gamma^2\right)^* \psi\right]\nonumber\\
    &- \frac{1}{4}\left(\psi^\dagger \gamma^2\gamma_5 \psi^*\right) \left[\psi^\text{T} \left(\gamma^2\gamma_5\right)^* \psi\right] +\nonumber\\
    &+\frac{1}{4}\left(\psi^\dagger \gamma^3 \psi \right) \left[\psi^\text{T} \left(\gamma^3\right)^* \psi\right]\nonumber\\
    &+\frac{1}{4}\left(\psi^\dagger \gamma^0\gamma_5 \psi \right) \left[\psi^\text{T} \left(\gamma^0\gamma_5\right)^* \psi\right] +\nonumber\\
    &+ \frac{1}{4}\left(\psi^\dagger \gamma^1 \psi \right) \left[\psi^\text{T} \left(\gamma^1\right)^* \psi\right] \nonumber\\
    &- \frac{1}{4}\left(\psi^\dagger \gamma^1\gamma^3 \psi \right) \left[\psi^\text{T} \left(\gamma^1\gamma^3\right)^* \psi\right] \nonumber\\
    &= \frac{1}{4}\sum_a \xi_a\left(\psi^\dagger O^a \psi \right) \left[\psi^\text{T} \left(O^a\right)^* \psi\right],
\end{align}
with $\xi_a = 1$ for $\{\gamma^0 \gamma^2, \, \gamma^3,\, \gamma^0 \gamma^5,\, \gamma^1\}$ and $\xi_a = -1$ for $\gamma^2 \gamma_5$ and $\gamma^1 \gamma^3$.

\bibliography{refs.bib}

\begin{thebibliography}{61}%
\makeatletter
\providecommand \@ifxundefined [1]{%
 \@ifx{#1\undefined}
}%
\providecommand \@ifnum [1]{%
 \ifnum #1\expandafter \@firstoftwo
 \else \expandafter \@secondoftwo
 \fi
}%
\providecommand \@ifx [1]{%
 \ifx #1\expandafter \@firstoftwo
 \else \expandafter \@secondoftwo
 \fi
}%
\providecommand \natexlab [1]{#1}%
\providecommand \enquote  [1]{``#1''}%
\providecommand \bibnamefont  [1]{#1}%
\providecommand \bibfnamefont [1]{#1}%
\providecommand \citenamefont [1]{#1}%
\providecommand \href@noop [0]{\@secondoftwo}%
\providecommand \href [0]{\begingroup \@sanitize@url \@href}%
\providecommand \@href[1]{\@@startlink{#1}\@@href}%
\providecommand \@@href[1]{\endgroup#1\@@endlink}%
\providecommand \@sanitize@url [0]{\catcode `\\12\catcode `\$12\catcode
  `\&12\catcode `\#12\catcode `\^12\catcode `\_12\catcode `\%12\relax}%
\providecommand \@@startlink[1]{}%
\providecommand \@@endlink[0]{}%
\providecommand \url  [0]{\begingroup\@sanitize@url \@url }%
\providecommand \@url [1]{\endgroup\@href {#1}{\urlprefix }}%
\providecommand \urlprefix  [0]{URL }%
\providecommand \Eprint [0]{\href }%
\providecommand \doibase [0]{https://doi.org/}%
\providecommand \selectlanguage [0]{\@gobble}%
\providecommand \bibinfo  [0]{\@secondoftwo}%
\providecommand \bibfield  [0]{\@secondoftwo}%
\providecommand \translation [1]{[#1]}%
\providecommand \BibitemOpen [0]{}%
\providecommand \bibitemStop [0]{}%
\providecommand \bibitemNoStop [0]{.\EOS\space}%
\providecommand \EOS [0]{\spacefactor3000\relax}%
\providecommand \BibitemShut  [1]{\csname bibitem#1\endcsname}%
\let\auto@bib@innerbib\@empty
\bibitem [{\citenamefont {Bertone}\ \emph {et~al.}(2005)\citenamefont
  {Bertone}, \citenamefont {Hooper},\ and\ \citenamefont
  {Silk}}]{Bertone:2004pz}%
  \BibitemOpen
  \bibfield  {author} {\bibinfo {author} {\bibfnamefont {G.}~\bibnamefont
  {Bertone}}, \bibinfo {author} {\bibfnamefont {D.}~\bibnamefont {Hooper}},\
  and\ \bibinfo {author} {\bibfnamefont {J.}~\bibnamefont {Silk}},\ }\bibfield
  {title} {\bibinfo {title} {{Particle dark matter: Evidence, candidates and
  constraints}},\ }\href {https://doi.org/10.1016/j.physrep.2004.08.031}
  {\bibfield  {journal} {\bibinfo  {journal} {Phys. Rept.}\ }\textbf {\bibinfo
  {volume} {405}},\ \bibinfo {pages} {279} (\bibinfo {year} {2005})},\ \Eprint
  {https://arxiv.org/abs/hep-ph/0404175} {arXiv:hep-ph/0404175} \BibitemShut
  {NoStop}%
\bibitem [{\citenamefont {Arbey}\ and\ \citenamefont
  {Mahmoudi}(2021)}]{Arbey:2021gdg}%
  \BibitemOpen
  \bibfield  {author} {\bibinfo {author} {\bibfnamefont {A.}~\bibnamefont
  {Arbey}}\ and\ \bibinfo {author} {\bibfnamefont {F.}~\bibnamefont
  {Mahmoudi}},\ }\bibfield  {title} {\bibinfo {title} {{Dark matter and the
  early Universe: a review}},\ }\href
  {https://doi.org/10.1016/j.ppnp.2021.103865} {\bibfield  {journal} {\bibinfo
  {journal} {Prog. Part. Nucl. Phys.}\ }\textbf {\bibinfo {volume} {119}},\
  \bibinfo {pages} {103865} (\bibinfo {year} {2021})},\ \Eprint
  {https://arxiv.org/abs/2104.11488} {arXiv:2104.11488 [hep-ph]} \BibitemShut
  {NoStop}%
\bibitem [{\citenamefont {Alexander}\ and\ \citenamefont {Cormack}(2017)}]{AC}%
  \BibitemOpen
  \bibfield  {author} {\bibinfo {author} {\bibfnamefont {S.}~\bibnamefont
  {Alexander}}\ and\ \bibinfo {author} {\bibfnamefont {S.}~\bibnamefont
  {Cormack}},\ }\bibfield  {title} {\bibinfo {title} {{Gravitationally bound
  BCS state as dark matter}},\ }\href
  {https://doi.org/10.1088/1475-7516/2017/04/005} {\bibfield  {journal}
  {\bibinfo  {journal} {JCAP}\ }\textbf {\bibinfo {volume} {04}},\ \bibinfo
  {pages} {005}},\ \Eprint {https://arxiv.org/abs/1607.08621} {arXiv:1607.08621
  [astro-ph.CO]} \BibitemShut {NoStop}%
\bibitem [{\citenamefont {Alexander}\ \emph {et~al.}(2018)\citenamefont
  {Alexander}, \citenamefont {McDonough},\ and\ \citenamefont
  {Spergel}}]{Alexander:2018fjp}%
  \BibitemOpen
  \bibfield  {author} {\bibinfo {author} {\bibfnamefont {S.}~\bibnamefont
  {Alexander}}, \bibinfo {author} {\bibfnamefont {E.}~\bibnamefont
  {McDonough}},\ and\ \bibinfo {author} {\bibfnamefont {D.~N.}\ \bibnamefont
  {Spergel}},\ }\bibfield  {title} {\bibinfo {title} {{Chiral Gravitational
  Waves and Baryon Superfluid Dark Matter}},\ }\href
  {https://doi.org/10.1088/1475-7516/2018/05/003} {\bibfield  {journal}
  {\bibinfo  {journal} {JCAP}\ }\textbf {\bibinfo {volume} {05}},\ \bibinfo
  {pages} {003}},\ \Eprint {https://arxiv.org/abs/1801.07255} {arXiv:1801.07255
  [hep-th]} \BibitemShut {NoStop}%
\bibitem [{\citenamefont {Garani}\ \emph {et~al.}(2022)\citenamefont {Garani},
  \citenamefont {Tytgat},\ and\ \citenamefont {Vandecasteele}}]{Garani}%
  \BibitemOpen
  \bibfield  {author} {\bibinfo {author} {\bibfnamefont {R.}~\bibnamefont
  {Garani}}, \bibinfo {author} {\bibfnamefont {M.~H.~G.}\ \bibnamefont
  {Tytgat}},\ and\ \bibinfo {author} {\bibfnamefont {J.}~\bibnamefont
  {Vandecasteele}},\ }\bibfield  {title} {\bibinfo {title} {{Condensed dark
  matter with a Yukawa interaction}},\ }\href
  {https://doi.org/10.1103/PhysRevD.106.116003} {\bibfield  {journal} {\bibinfo
   {journal} {Phys. Rev. D}\ }\textbf {\bibinfo {volume} {106}},\ \bibinfo
  {pages} {116003} (\bibinfo {year} {2022})},\ \Eprint
  {https://arxiv.org/abs/2207.06928} {arXiv:2207.06928 [hep-ph]} \BibitemShut
  {NoStop}%
\bibitem [{\citenamefont {Tong}\ \emph {et~al.}(2024)\citenamefont {Tong},
  \citenamefont {Wang}, \citenamefont {Zhang},\ and\ \citenamefont
  {Zhu}}]{Tong:2023krn}%
  \BibitemOpen
  \bibfield  {author} {\bibinfo {author} {\bibfnamefont {X.}~\bibnamefont
  {Tong}}, \bibinfo {author} {\bibfnamefont {Y.}~\bibnamefont {Wang}}, \bibinfo
  {author} {\bibfnamefont {C.}~\bibnamefont {Zhang}},\ and\ \bibinfo {author}
  {\bibfnamefont {Y.}~\bibnamefont {Zhu}},\ }\bibfield  {title} {\bibinfo
  {title} {{BCS in the sky: signatures of inflationary fermion condensation}},\
  }\href {https://doi.org/10.1088/1475-7516/2024/04/022} {\bibfield  {journal}
  {\bibinfo  {journal} {JCAP}\ }\textbf {\bibinfo {volume} {04}},\ \bibinfo
  {pages} {022}},\ \Eprint {https://arxiv.org/abs/2304.09428} {arXiv:2304.09428
  [hep-th]} \BibitemShut {NoStop}%
\bibitem [{\citenamefont {Spergel}\ \emph {et~al.}(2007)\citenamefont {Spergel}
  \emph {et~al.}}]{WMAP:2006bqn}%
  \BibitemOpen
  \bibfield  {author} {\bibinfo {author} {\bibfnamefont {D.~N.}\ \bibnamefont
  {Spergel}} \emph {et~al.} (\bibinfo {collaboration} {WMAP}),\ }\bibfield
  {title} {\bibinfo {title} {{Wilkinson Microwave Anisotropy Probe (WMAP) three
  year results: implications for cosmology}},\ }\href
  {https://doi.org/10.1086/513700} {\bibfield  {journal} {\bibinfo  {journal}
  {Astrophys. J. Suppl.}\ }\textbf {\bibinfo {volume} {170}},\ \bibinfo {pages}
  {377} (\bibinfo {year} {2007})},\ \Eprint
  {https://arxiv.org/abs/astro-ph/0603449} {arXiv:astro-ph/0603449}
  \BibitemShut {NoStop}%
\bibitem [{\citenamefont {Aghanim}\ \emph {et~al.}(2020)\citenamefont {Aghanim}
  \emph {et~al.}}]{Planck:2018vyg}%
  \BibitemOpen
  \bibfield  {author} {\bibinfo {author} {\bibfnamefont {N.}~\bibnamefont
  {Aghanim}} \emph {et~al.} (\bibinfo {collaboration} {Planck}),\ }\bibfield
  {title} {\bibinfo {title} {{Planck 2018 results. VI. Cosmological
  parameters}},\ }\href {https://doi.org/10.1051/0004-6361/201833910}
  {\bibfield  {journal} {\bibinfo  {journal} {Astron. Astrophys.}\ }\textbf
  {\bibinfo {volume} {641}},\ \bibinfo {pages} {A6} (\bibinfo {year} {2020})},\
  \bibinfo {note} {[Erratum: Astron.Astrophys. 652, C4 (2021)]},\ \Eprint
  {https://arxiv.org/abs/1807.06209} {arXiv:1807.06209 [astro-ph.CO]}
  \BibitemShut {NoStop}%
\bibitem [{\citenamefont {McDonald}(2011{\natexlab{a}})}]{McDonald:2010toz}%
  \BibitemOpen
  \bibfield  {author} {\bibinfo {author} {\bibfnamefont {J.}~\bibnamefont
  {McDonald}},\ }\bibfield  {title} {\bibinfo {title} {{Baryomorphosis:
  Relating the Baryon Asymmetry to the 'WIMP Miracle'}},\ }\href
  {https://doi.org/10.1103/PhysRevD.83.083509} {\bibfield  {journal} {\bibinfo
  {journal} {Phys. Rev. D}\ }\textbf {\bibinfo {volume} {83}},\ \bibinfo
  {pages} {083509} (\bibinfo {year} {2011}{\natexlab{a}})},\ \Eprint
  {https://arxiv.org/abs/1009.3227} {arXiv:1009.3227 [hep-ph]} \BibitemShut
  {NoStop}%
\bibitem [{\citenamefont {Davoudiasl}\ and\ \citenamefont
  {Mohapatra}(2012)}]{Davoudiasl:2012uw}%
  \BibitemOpen
  \bibfield  {author} {\bibinfo {author} {\bibfnamefont {H.}~\bibnamefont
  {Davoudiasl}}\ and\ \bibinfo {author} {\bibfnamefont {R.~N.}\ \bibnamefont
  {Mohapatra}},\ }\bibfield  {title} {\bibinfo {title} {{On Relating the
  Genesis of Cosmic Baryons and Dark Matter}},\ }\href
  {https://doi.org/10.1088/1367-2630/14/9/095011} {\bibfield  {journal}
  {\bibinfo  {journal} {New J. Phys.}\ }\textbf {\bibinfo {volume} {14}},\
  \bibinfo {pages} {095011} (\bibinfo {year} {2012})},\ \Eprint
  {https://arxiv.org/abs/1203.1247} {arXiv:1203.1247 [hep-ph]} \BibitemShut
  {NoStop}%
\bibitem [{\citenamefont {Petraki}\ and\ \citenamefont
  {Volkas}(2013)}]{Petraki:2013wwa}%
  \BibitemOpen
  \bibfield  {author} {\bibinfo {author} {\bibfnamefont {K.}~\bibnamefont
  {Petraki}}\ and\ \bibinfo {author} {\bibfnamefont {R.~R.}\ \bibnamefont
  {Volkas}},\ }\bibfield  {title} {\bibinfo {title} {{Review of asymmetric dark
  matter}},\ }\href {https://doi.org/10.1142/S0217751X13300287} {\bibfield
  {journal} {\bibinfo  {journal} {Int. J. Mod. Phys. A}\ }\textbf {\bibinfo
  {volume} {28}},\ \bibinfo {pages} {1330028} (\bibinfo {year} {2013})},\
  \Eprint {https://arxiv.org/abs/1305.4939} {arXiv:1305.4939 [hep-ph]}
  \BibitemShut {NoStop}%
\bibitem [{\citenamefont {Arcadi}\ \emph {et~al.}(2018)\citenamefont {Arcadi},
  \citenamefont {Dutra}, \citenamefont {Ghosh}, \citenamefont {Lindner},
  \citenamefont {Mambrini}, \citenamefont {Pierre}, \citenamefont {Profumo},\
  and\ \citenamefont {Queiroz}}]{Arcadi:2017kky}%
  \BibitemOpen
  \bibfield  {author} {\bibinfo {author} {\bibfnamefont {G.}~\bibnamefont
  {Arcadi}}, \bibinfo {author} {\bibfnamefont {M.}~\bibnamefont {Dutra}},
  \bibinfo {author} {\bibfnamefont {P.}~\bibnamefont {Ghosh}}, \bibinfo
  {author} {\bibfnamefont {M.}~\bibnamefont {Lindner}}, \bibinfo {author}
  {\bibfnamefont {Y.}~\bibnamefont {Mambrini}}, \bibinfo {author}
  {\bibfnamefont {M.}~\bibnamefont {Pierre}}, \bibinfo {author} {\bibfnamefont
  {S.}~\bibnamefont {Profumo}},\ and\ \bibinfo {author} {\bibfnamefont {F.~S.}\
  \bibnamefont {Queiroz}},\ }\bibfield  {title} {\bibinfo {title} {{The waning
  of the WIMP? A review of models, searches, and constraints}},\ }\href
  {https://doi.org/10.1140/epjc/s10052-018-5662-y} {\bibfield  {journal}
  {\bibinfo  {journal} {Eur. Phys. J. C}\ }\textbf {\bibinfo {volume} {78}},\
  \bibinfo {pages} {203} (\bibinfo {year} {2018})},\ \Eprint
  {https://arxiv.org/abs/1703.07364} {arXiv:1703.07364 [hep-ph]} \BibitemShut
  {NoStop}%
\bibitem [{\citenamefont {McDonald}(2011{\natexlab{b}})}]{McDonald:2011sv}%
  \BibitemOpen
  \bibfield  {author} {\bibinfo {author} {\bibfnamefont {J.}~\bibnamefont
  {McDonald}},\ }\bibfield  {title} {\bibinfo {title} {{Simultaneous Generation
  of WIMP Miracle-like Densities of Baryons and Dark Matter}},\ }\href
  {https://doi.org/10.1103/PhysRevD.84.103514} {\bibfield  {journal} {\bibinfo
  {journal} {Phys. Rev. D}\ }\textbf {\bibinfo {volume} {84}},\ \bibinfo
  {pages} {103514} (\bibinfo {year} {2011}{\natexlab{b}})},\ \Eprint
  {https://arxiv.org/abs/1108.4653} {arXiv:1108.4653 [hep-ph]} \BibitemShut
  {NoStop}%
\bibitem [{\citenamefont {Cui}\ \emph {et~al.}(2012)\citenamefont {Cui},
  \citenamefont {Randall},\ and\ \citenamefont {Shuve}}]{Cui:2011ab}%
  \BibitemOpen
  \bibfield  {author} {\bibinfo {author} {\bibfnamefont {Y.}~\bibnamefont
  {Cui}}, \bibinfo {author} {\bibfnamefont {L.}~\bibnamefont {Randall}},\ and\
  \bibinfo {author} {\bibfnamefont {B.}~\bibnamefont {Shuve}},\ }\bibfield
  {title} {\bibinfo {title} {{A WIMPy Baryogenesis Miracle}},\ }\href
  {https://doi.org/10.1007/JHEP04(2012)075} {\bibfield  {journal} {\bibinfo
  {journal} {JHEP}\ }\textbf {\bibinfo {volume} {04}},\ \bibinfo {pages}
  {075}},\ \Eprint {https://arxiv.org/abs/1112.2704} {arXiv:1112.2704 [hep-ph]}
  \BibitemShut {NoStop}%
\bibitem [{\citenamefont {Salucci}(2019)}]{Salucci:2018hqu}%
  \BibitemOpen
  \bibfield  {author} {\bibinfo {author} {\bibfnamefont {P.}~\bibnamefont
  {Salucci}},\ }\bibfield  {title} {\bibinfo {title} {{The distribution of dark
  matter in galaxies}},\ }\href {https://doi.org/10.1007/s00159-018-0113-1}
  {\bibfield  {journal} {\bibinfo  {journal} {Astron. Astrophys. Rev.}\
  }\textbf {\bibinfo {volume} {27}},\ \bibinfo {pages} {2} (\bibinfo {year}
  {2019})},\ \Eprint {https://arxiv.org/abs/1811.08843} {arXiv:1811.08843
  [astro-ph.GA]} \BibitemShut {NoStop}%
\bibitem [{\citenamefont {Witten}(1984)}]{Witten:1984rs}%
  \BibitemOpen
  \bibfield  {author} {\bibinfo {author} {\bibfnamefont {E.}~\bibnamefont
  {Witten}},\ }\bibfield  {title} {\bibinfo {title} {{Cosmic Separation of
  Phases}},\ }\href {https://doi.org/10.1103/PhysRevD.30.272} {\bibfield
  {journal} {\bibinfo  {journal} {Phys. Rev. D}\ }\textbf {\bibinfo {volume}
  {30}},\ \bibinfo {pages} {272} (\bibinfo {year} {1984})}\BibitemShut
  {NoStop}%
\bibitem [{\citenamefont {Farhi}\ and\ \citenamefont
  {Jaffe}(1984)}]{Farhi:1984qu}%
  \BibitemOpen
  \bibfield  {author} {\bibinfo {author} {\bibfnamefont {E.}~\bibnamefont
  {Farhi}}\ and\ \bibinfo {author} {\bibfnamefont {R.~L.}\ \bibnamefont
  {Jaffe}},\ }\bibfield  {title} {\bibinfo {title} {{Strange Matter}},\ }\href
  {https://doi.org/10.1103/PhysRevD.30.2379} {\bibfield  {journal} {\bibinfo
  {journal} {Phys. Rev. D}\ }\textbf {\bibinfo {volume} {30}},\ \bibinfo
  {pages} {2379} (\bibinfo {year} {1984})}\BibitemShut {NoStop}%
\bibitem [{\citenamefont {Zhitnitsky}(2003)}]{Zhitnitsky:2002qa}%
  \BibitemOpen
  \bibfield  {author} {\bibinfo {author} {\bibfnamefont {A.~R.}\ \bibnamefont
  {Zhitnitsky}},\ }\bibfield  {title} {\bibinfo {title} {{'Nonbaryonic' dark
  matter as baryonic color superconductor}},\ }\href
  {https://doi.org/10.1088/1475-7516/2003/10/010} {\bibfield  {journal}
  {\bibinfo  {journal} {JCAP}\ }\textbf {\bibinfo {volume} {10}},\ \bibinfo
  {pages} {010}},\ \Eprint {https://arxiv.org/abs/hep-ph/0202161}
  {arXiv:hep-ph/0202161} \BibitemShut {NoStop}%
\bibitem [{\citenamefont {Cooper}(1956)}]{PhysRev.104.1189}%
  \BibitemOpen
  \bibfield  {author} {\bibinfo {author} {\bibfnamefont {L.~N.}\ \bibnamefont
  {Cooper}},\ }\bibfield  {title} {\bibinfo {title} {Bound electron pairs in a
  degenerate fermi gas},\ }\href {https://doi.org/10.1103/PhysRev.104.1189}
  {\bibfield  {journal} {\bibinfo  {journal} {Phys. Rev.}\ }\textbf {\bibinfo
  {volume} {104}},\ \bibinfo {pages} {1189} (\bibinfo {year}
  {1956})}\BibitemShut {NoStop}%
\bibitem [{\citenamefont {Bardeen}\ \emph
  {et~al.}(1957{\natexlab{a}})\citenamefont {Bardeen}, \citenamefont {Cooper},\
  and\ \citenamefont {Schrieffer}}]{PhysRev.106.162}%
  \BibitemOpen
  \bibfield  {author} {\bibinfo {author} {\bibfnamefont {J.}~\bibnamefont
  {Bardeen}}, \bibinfo {author} {\bibfnamefont {L.~N.}\ \bibnamefont
  {Cooper}},\ and\ \bibinfo {author} {\bibfnamefont {J.~R.}\ \bibnamefont
  {Schrieffer}},\ }\bibfield  {title} {\bibinfo {title} {Microscopic theory of
  superconductivity},\ }\href {https://doi.org/10.1103/PhysRev.106.162}
  {\bibfield  {journal} {\bibinfo  {journal} {Phys. Rev.}\ }\textbf {\bibinfo
  {volume} {106}},\ \bibinfo {pages} {162} (\bibinfo {year}
  {1957}{\natexlab{a}})}\BibitemShut {NoStop}%
\bibitem [{\citenamefont {Bardeen}\ \emph
  {et~al.}(1957{\natexlab{b}})\citenamefont {Bardeen}, \citenamefont {Cooper},\
  and\ \citenamefont {Schrieffer}}]{PhysRev.108.1175}%
  \BibitemOpen
  \bibfield  {author} {\bibinfo {author} {\bibfnamefont {J.}~\bibnamefont
  {Bardeen}}, \bibinfo {author} {\bibfnamefont {L.~N.}\ \bibnamefont
  {Cooper}},\ and\ \bibinfo {author} {\bibfnamefont {J.~R.}\ \bibnamefont
  {Schrieffer}},\ }\bibfield  {title} {\bibinfo {title} {Theory of
  superconductivity},\ }\href {https://doi.org/10.1103/PhysRev.108.1175}
  {\bibfield  {journal} {\bibinfo  {journal} {Phys. Rev.}\ }\textbf {\bibinfo
  {volume} {108}},\ \bibinfo {pages} {1175} (\bibinfo {year}
  {1957}{\natexlab{b}})}\BibitemShut {NoStop}%
\bibitem [{\citenamefont {Alexander}\ \emph {et~al.}(2021)\citenamefont
  {Alexander}, \citenamefont {McDonough},\ and\ \citenamefont
  {Spergel}}]{Alexander:2020wpm}%
  \BibitemOpen
  \bibfield  {author} {\bibinfo {author} {\bibfnamefont {S.}~\bibnamefont
  {Alexander}}, \bibinfo {author} {\bibfnamefont {E.}~\bibnamefont
  {McDonough}},\ and\ \bibinfo {author} {\bibfnamefont {D.~N.}\ \bibnamefont
  {Spergel}},\ }\bibfield  {title} {\bibinfo {title} {{Strongly-interacting
  ultralight millicharged particles}},\ }\href
  {https://doi.org/10.1016/j.physletb.2021.136653} {\bibfield  {journal}
  {\bibinfo  {journal} {Phys. Lett. B}\ }\textbf {\bibinfo {volume} {822}},\
  \bibinfo {pages} {136653} (\bibinfo {year} {2021})},\ \Eprint
  {https://arxiv.org/abs/2011.06589} {arXiv:2011.06589 [astro-ph.CO]}
  \BibitemShut {NoStop}%
\bibitem [{\citenamefont {Nambu}\ and\ \citenamefont
  {Jona-Lasinio}(1961{\natexlab{a}})}]{Nambu:1961tp}%
  \BibitemOpen
  \bibfield  {author} {\bibinfo {author} {\bibfnamefont {Y.}~\bibnamefont
  {Nambu}}\ and\ \bibinfo {author} {\bibfnamefont {G.}~\bibnamefont
  {Jona-Lasinio}},\ }\bibfield  {title} {\bibinfo {title} {{Dynamical Model of
  Elementary Particles Based on an Analogy with Superconductivity. 1.}},\
  }\href {https://doi.org/10.1103/PhysRev.122.345} {\bibfield  {journal}
  {\bibinfo  {journal} {Phys. Rev.}\ }\textbf {\bibinfo {volume} {122}},\
  \bibinfo {pages} {345} (\bibinfo {year} {1961}{\natexlab{a}})}\BibitemShut
  {NoStop}%
\bibitem [{\citenamefont {Nambu}\ and\ \citenamefont
  {Jona-Lasinio}(1961{\natexlab{b}})}]{Nambu:1961fr}%
  \BibitemOpen
  \bibfield  {author} {\bibinfo {author} {\bibfnamefont {Y.}~\bibnamefont
  {Nambu}}\ and\ \bibinfo {author} {\bibfnamefont {G.}~\bibnamefont
  {Jona-Lasinio}},\ }\bibfield  {title} {\bibinfo {title} {{Dynamical model of
  elementary particles based on an analogy with superconductivity. II.}},\
  }\href {https://doi.org/10.1103/PhysRev.124.246} {\bibfield  {journal}
  {\bibinfo  {journal} {Phys. Rev.}\ }\textbf {\bibinfo {volume} {124}},\
  \bibinfo {pages} {246} (\bibinfo {year} {1961}{\natexlab{b}})}\BibitemShut
  {NoStop}%
\bibitem [{\citenamefont {Stratonovich}(1957)}]{stratonovich1957method}%
  \BibitemOpen
  \bibfield  {author} {\bibinfo {author} {\bibfnamefont {R.}~\bibnamefont
  {Stratonovich}},\ }\bibfield  {title} {\bibinfo {title} {On a method of
  calculating quantum distribution functions},\ }in\ \href@noop {} {\emph
  {\bibinfo {booktitle} {Soviet Physics Doklady}}},\ Vol.~\bibinfo {volume}
  {2}\ (\bibinfo {year} {1957})\ p.\ \bibinfo {pages} {416}\BibitemShut
  {NoStop}%
\bibitem [{\citenamefont {Hubbard}(1959)}]{Hubbard:1959ub}%
  \BibitemOpen
  \bibfield  {author} {\bibinfo {author} {\bibfnamefont {J.}~\bibnamefont
  {Hubbard}},\ }\bibfield  {title} {\bibinfo {title} {{Calculation of partition
  functions}},\ }\href {https://doi.org/10.1103/PhysRevLett.3.77} {\bibfield
  {journal} {\bibinfo  {journal} {Phys. Rev. Lett.}\ }\textbf {\bibinfo
  {volume} {3}},\ \bibinfo {pages} {77} (\bibinfo {year} {1959})}\BibitemShut
  {NoStop}%
\bibitem [{\citenamefont {Gor'kov}(1958)}]{Gorkov1958}%
  \BibitemOpen
  \bibfield  {author} {\bibinfo {author} {\bibfnamefont {L.}~\bibnamefont
  {Gor'kov}},\ }\bibfield  {title} {\bibinfo {title} {{On the energy spectrum
  of superconductors}},\ }\href@noop {} {\bibfield  {journal} {\bibinfo
  {journal} {Sov. Phys. JETP}\ }\textbf {\bibinfo {volume} {7}},\ \bibinfo
  {pages} {505} (\bibinfo {year} {1958})}\BibitemShut {NoStop}%
\bibitem [{\citenamefont {Nambu}(1960)}]{Nambu:1960tm}%
  \BibitemOpen
  \bibfield  {author} {\bibinfo {author} {\bibfnamefont {Y.}~\bibnamefont
  {Nambu}},\ }\bibfield  {title} {\bibinfo {title} {{Quasiparticles and Gauge
  Invariance in the Theory of Superconductivity}},\ }\href
  {https://doi.org/10.1103/PhysRev.117.648} {\bibfield  {journal} {\bibinfo
  {journal} {Phys. Rev.}\ }\textbf {\bibinfo {volume} {117}},\ \bibinfo {pages}
  {648} (\bibinfo {year} {1960})}\BibitemShut {NoStop}%
\bibitem [{\citenamefont {Polchinski}(1992)}]{Joe}%
  \BibitemOpen
  \bibfield  {author} {\bibinfo {author} {\bibfnamefont {J.}~\bibnamefont
  {Polchinski}},\ }\bibfield  {title} {\bibinfo {title} {{Effective field
  theory and the Fermi surface}},\ }in\ \href@noop {} {\emph {\bibinfo
  {booktitle} {{Theoretical Advanced Study Institute (TASI 92): From Black
  Holes and Strings to Particles}}}}\ (\bibinfo {year} {1992})\ pp.\ \bibinfo
  {pages} {0235--276},\ \Eprint {https://arxiv.org/abs/hep-th/9210046}
  {arXiv:hep-th/9210046} \BibitemShut {NoStop}%
\bibitem [{\citenamefont {Shankar}(1994)}]{Shankar:1993pf}%
  \BibitemOpen
  \bibfield  {author} {\bibinfo {author} {\bibfnamefont {R.}~\bibnamefont
  {Shankar}},\ }\bibfield  {title} {\bibinfo {title} {{Renormalization group
  approach to interacting fermions}},\ }\href
  {https://doi.org/10.1103/RevModPhys.66.129} {\bibfield  {journal} {\bibinfo
  {journal} {Rev. Mod. Phys.}\ }\textbf {\bibinfo {volume} {66}},\ \bibinfo
  {pages} {129} (\bibinfo {year} {1994})},\ \Eprint
  {https://arxiv.org/abs/cond-mat/9307009} {arXiv:cond-mat/9307009}
  \BibitemShut {NoStop}%
\bibitem [{\citenamefont {Anglani}\ \emph {et~al.}(2011)\citenamefont
  {Anglani}, \citenamefont {Mannarelli},\ and\ \citenamefont
  {Ruggieri}}]{Anglani:2011cw}%
  \BibitemOpen
  \bibfield  {author} {\bibinfo {author} {\bibfnamefont {R.}~\bibnamefont
  {Anglani}}, \bibinfo {author} {\bibfnamefont {M.}~\bibnamefont
  {Mannarelli}},\ and\ \bibinfo {author} {\bibfnamefont {M.}~\bibnamefont
  {Ruggieri}},\ }\bibfield  {title} {\bibinfo {title} {{Collective modes in the
  color flavor locked phase}},\ }\href
  {https://doi.org/10.1088/1367-2630/13/5/055002} {\bibfield  {journal}
  {\bibinfo  {journal} {New J. Phys.}\ }\textbf {\bibinfo {volume} {13}},\
  \bibinfo {pages} {055002} (\bibinfo {year} {2011})},\ \Eprint
  {https://arxiv.org/abs/1101.4277} {arXiv:1101.4277 [hep-ph]} \BibitemShut
  {NoStop}%
\bibitem [{\citenamefont {Adshead}\ and\ \citenamefont
  {Sfakianakis}(2016)}]{Adshead:2015jza}%
  \BibitemOpen
  \bibfield  {author} {\bibinfo {author} {\bibfnamefont {P.}~\bibnamefont
  {Adshead}}\ and\ \bibinfo {author} {\bibfnamefont {E.~I.}\ \bibnamefont
  {Sfakianakis}},\ }\bibfield  {title} {\bibinfo {title} {{Leptogenesis from
  left-handed neutrino production during axion inflation}},\ }\href
  {https://doi.org/10.1103/PhysRevLett.116.091301} {\bibfield  {journal}
  {\bibinfo  {journal} {Phys. Rev. Lett.}\ }\textbf {\bibinfo {volume} {116}},\
  \bibinfo {pages} {091301} (\bibinfo {year} {2016})},\ \Eprint
  {https://arxiv.org/abs/1508.00881} {arXiv:1508.00881 [hep-ph]} \BibitemShut
  {NoStop}%
\bibitem [{\citenamefont {Sakharov}(1967)}]{Sakharov:1967dj}%
  \BibitemOpen
  \bibfield  {author} {\bibinfo {author} {\bibfnamefont {A.~D.}\ \bibnamefont
  {Sakharov}},\ }\bibfield  {title} {\bibinfo {title} {{Violation of CP
  Invariance, C asymmetry, and baryon asymmetry of the universe}},\ }\href
  {https://doi.org/10.1070/PU1991v034n05ABEH002497} {\bibfield  {journal}
  {\bibinfo  {journal} {Pisma Zh. Eksp. Teor. Fiz.}\ }\textbf {\bibinfo
  {volume} {5}},\ \bibinfo {pages} {32} (\bibinfo {year} {1967})}\BibitemShut
  {NoStop}%
\bibitem [{\citenamefont {Cohen}\ \emph {et~al.}(1993)\citenamefont {Cohen},
  \citenamefont {Kaplan},\ and\ \citenamefont {Nelson}}]{Cohen:1993nk}%
  \BibitemOpen
  \bibfield  {author} {\bibinfo {author} {\bibfnamefont {A.~G.}\ \bibnamefont
  {Cohen}}, \bibinfo {author} {\bibfnamefont {D.~B.}\ \bibnamefont {Kaplan}},\
  and\ \bibinfo {author} {\bibfnamefont {A.~E.}\ \bibnamefont {Nelson}},\
  }\bibfield  {title} {\bibinfo {title} {{Progress in electroweak
  baryogenesis}},\ }\href {https://doi.org/10.1146/annurev.ns.43.120193.000331}
  {\bibfield  {journal} {\bibinfo  {journal} {Ann. Rev. Nucl. Part. Sci.}\
  }\textbf {\bibinfo {volume} {43}},\ \bibinfo {pages} {27} (\bibinfo {year}
  {1993})},\ \Eprint {https://arxiv.org/abs/hep-ph/9302210}
  {arXiv:hep-ph/9302210} \BibitemShut {NoStop}%
\bibitem [{\citenamefont {Riotto}(1998)}]{Riotto:1998bt}%
  \BibitemOpen
  \bibfield  {author} {\bibinfo {author} {\bibfnamefont {A.}~\bibnamefont
  {Riotto}},\ }\bibfield  {title} {\bibinfo {title} {{Theories of
  baryogenesis}},\ }in\ \href@noop {} {\emph {\bibinfo {booktitle} {{ICTP
  Summer School in High-Energy Physics and Cosmology}}}}\ (\bibinfo {year}
  {1998})\ pp.\ \bibinfo {pages} {326--436},\ \Eprint
  {https://arxiv.org/abs/hep-ph/9807454} {arXiv:hep-ph/9807454} \BibitemShut
  {NoStop}%
\bibitem [{\citenamefont {Trodden}(1999)}]{Trodden:1998ym}%
  \BibitemOpen
  \bibfield  {author} {\bibinfo {author} {\bibfnamefont {M.}~\bibnamefont
  {Trodden}},\ }\bibfield  {title} {\bibinfo {title} {{Electroweak
  baryogenesis}},\ }\href {https://doi.org/10.1103/RevModPhys.71.1463}
  {\bibfield  {journal} {\bibinfo  {journal} {Rev. Mod. Phys.}\ }\textbf
  {\bibinfo {volume} {71}},\ \bibinfo {pages} {1463} (\bibinfo {year}
  {1999})},\ \Eprint {https://arxiv.org/abs/hep-ph/9803479}
  {arXiv:hep-ph/9803479} \BibitemShut {NoStop}%
\bibitem [{\citenamefont {Morrissey}\ and\ \citenamefont
  {Ramsey-Musolf}(2012)}]{Morrissey:2012db}%
  \BibitemOpen
  \bibfield  {author} {\bibinfo {author} {\bibfnamefont {D.~E.}\ \bibnamefont
  {Morrissey}}\ and\ \bibinfo {author} {\bibfnamefont {M.~J.}\ \bibnamefont
  {Ramsey-Musolf}},\ }\bibfield  {title} {\bibinfo {title} {{Electroweak
  baryogenesis}},\ }\href {https://doi.org/10.1088/1367-2630/14/12/125003}
  {\bibfield  {journal} {\bibinfo  {journal} {New J. Phys.}\ }\textbf {\bibinfo
  {volume} {14}},\ \bibinfo {pages} {125003} (\bibinfo {year} {2012})},\
  \Eprint {https://arxiv.org/abs/1206.2942} {arXiv:1206.2942 [hep-ph]}
  \BibitemShut {NoStop}%
\bibitem [{Note1()}]{Note1}%
  \BibitemOpen
  \bibinfo {note} {By ``sterile'' we mean that the right-handed neutrinos are
  singlets of the SM gauge groups. Strictly speaking, in the neutrino case we
  have a superfluid condensate, but our formalism generally accommodates
  superconductor states.}\BibitemShut {Stop}%
\bibitem [{\citenamefont {Kapusta}(2004)}]{Kapusta:2004gi}%
  \BibitemOpen
  \bibfield  {author} {\bibinfo {author} {\bibfnamefont {J.~I.}\ \bibnamefont
  {Kapusta}},\ }\bibfield  {title} {\bibinfo {title} {{Neutrino
  superfluidity}},\ }\href {https://doi.org/10.1103/PhysRevLett.93.251801}
  {\bibfield  {journal} {\bibinfo  {journal} {Phys. Rev. Lett.}\ }\textbf
  {\bibinfo {volume} {93}},\ \bibinfo {pages} {251801} (\bibinfo {year}
  {2004})},\ \Eprint {https://arxiv.org/abs/hep-th/0407164}
  {arXiv:hep-th/0407164} \BibitemShut {NoStop}%
\bibitem [{\citenamefont {Pisarski}\ and\ \citenamefont
  {Rischke}(1999)}]{Pisarski:1999av}%
  \BibitemOpen
  \bibfield  {author} {\bibinfo {author} {\bibfnamefont {R.~D.}\ \bibnamefont
  {Pisarski}}\ and\ \bibinfo {author} {\bibfnamefont {D.~H.}\ \bibnamefont
  {Rischke}},\ }\bibfield  {title} {\bibinfo {title} {{Superfluidity in a model
  of massless fermions coupled to scalar bosons}},\ }\href
  {https://doi.org/10.1103/PhysRevD.60.094013} {\bibfield  {journal} {\bibinfo
  {journal} {Phys. Rev. D}\ }\textbf {\bibinfo {volume} {60}},\ \bibinfo
  {pages} {094013} (\bibinfo {year} {1999})},\ \Eprint
  {https://arxiv.org/abs/nucl-th/9903023} {arXiv:nucl-th/9903023} \BibitemShut
  {NoStop}%
\bibitem [{\citenamefont {Chodos}\ and\ \citenamefont
  {Cooper}(2020)}]{Chodos:2020dgi}%
  \BibitemOpen
  \bibfield  {author} {\bibinfo {author} {\bibfnamefont {A.}~\bibnamefont
  {Chodos}}\ and\ \bibinfo {author} {\bibfnamefont {F.}~\bibnamefont
  {Cooper}},\ }\bibfield  {title} {\bibinfo {title} {{Neutrino condensation
  from a New Higgs Interaction}},\ }\href
  {https://doi.org/10.1103/PhysRevD.102.113003} {\bibfield  {journal} {\bibinfo
   {journal} {Phys. Rev. D}\ }\textbf {\bibinfo {volume} {102}},\ \bibinfo
  {pages} {113003} (\bibinfo {year} {2020})},\ \Eprint
  {https://arxiv.org/abs/2004.03731} {arXiv:2004.03731 [hep-ph]} \BibitemShut
  {NoStop}%
\bibitem [{\citenamefont {Arg\"uelles}\ \emph {et~al.}(2023)\citenamefont
  {Arg\"uelles}, \citenamefont {Becerra-Vergara}, \citenamefont {Rueda},\ and\
  \citenamefont {Ruffini}}]{Arguelles:2023nlh}%
  \BibitemOpen
  \bibfield  {author} {\bibinfo {author} {\bibfnamefont {C.~R.}\ \bibnamefont
  {Arg\"uelles}}, \bibinfo {author} {\bibfnamefont {E.~A.}\ \bibnamefont
  {Becerra-Vergara}}, \bibinfo {author} {\bibfnamefont {J.~A.}\ \bibnamefont
  {Rueda}},\ and\ \bibinfo {author} {\bibfnamefont {R.}~\bibnamefont
  {Ruffini}},\ }\bibfield  {title} {\bibinfo {title} {{Fermionic Dark Matter:
  Physics, Astrophysics, and Cosmology}},\ }\href
  {https://doi.org/10.3390/universe9040197} {\bibfield  {journal} {\bibinfo
  {journal} {Universe}\ }\textbf {\bibinfo {volume} {9}},\ \bibinfo {pages}
  {197} (\bibinfo {year} {2023})},\ \Eprint {https://arxiv.org/abs/2304.06329}
  {arXiv:2304.06329 [astro-ph.GA]} \BibitemShut {NoStop}%
\bibitem [{\citenamefont {Tremaine}\ and\ \citenamefont
  {Gunn}(1979)}]{PhysRevLett.42.407}%
  \BibitemOpen
  \bibfield  {author} {\bibinfo {author} {\bibfnamefont {S.}~\bibnamefont
  {Tremaine}}\ and\ \bibinfo {author} {\bibfnamefont {J.~E.}\ \bibnamefont
  {Gunn}},\ }\bibfield  {title} {\bibinfo {title} {Dynamical role of light
  neutral leptons in cosmology},\ }\href
  {https://doi.org/10.1103/PhysRevLett.42.407} {\bibfield  {journal} {\bibinfo
  {journal} {Phys. Rev. Lett.}\ }\textbf {\bibinfo {volume} {42}},\ \bibinfo
  {pages} {407} (\bibinfo {year} {1979})}\BibitemShut {NoStop}%
\bibitem [{\citenamefont {Workman}\ \emph {et~al.}(2022)\citenamefont {Workman}
  \emph {et~al.}}]{Workman:2022ynf}%
  \BibitemOpen
  \bibfield  {author} {\bibinfo {author} {\bibfnamefont {R.~L.}\ \bibnamefont
  {Workman}} \emph {et~al.} (\bibinfo {collaboration} {Particle Data Group}),\
  }\bibfield  {title} {\bibinfo {title} {{Review of Particle Physics}},\ }\href
  {https://doi.org/10.1093/ptep/ptac097} {\bibfield  {journal} {\bibinfo
  {journal} {PTEP}\ }\textbf {\bibinfo {volume} {2022}},\ \bibinfo {pages}
  {083C01} (\bibinfo {year} {2022})}\BibitemShut {NoStop}%
\bibitem [{\citenamefont {Trautner}(2017)}]{Trautner:2016ias}%
  \BibitemOpen
  \bibfield  {author} {\bibinfo {author} {\bibfnamefont {A.}~\bibnamefont
  {Trautner}},\ }\bibfield  {title} {\bibinfo {title} {{Massive Fermi Gas in
  the Expanding Universe}},\ }\href
  {https://doi.org/10.1088/1475-7516/2017/03/019} {\bibfield  {journal}
  {\bibinfo  {journal} {JCAP}\ }\textbf {\bibinfo {volume} {03}},\ \bibinfo
  {pages} {019}},\ \Eprint {https://arxiv.org/abs/1612.07249} {arXiv:1612.07249
  [astro-ph.CO]} \BibitemShut {NoStop}%
\bibitem [{\citenamefont {Chen}\ \emph {et~al.}(2015)\citenamefont {Chen},
  \citenamefont {Ratz},\ and\ \citenamefont {Trautner}}]{Chen:2015dka}%
  \BibitemOpen
  \bibfield  {author} {\bibinfo {author} {\bibfnamefont {M.-C.}\ \bibnamefont
  {Chen}}, \bibinfo {author} {\bibfnamefont {M.}~\bibnamefont {Ratz}},\ and\
  \bibinfo {author} {\bibfnamefont {A.}~\bibnamefont {Trautner}},\ }\bibfield
  {title} {\bibinfo {title} {{Nonthermal cosmic neutrino background}},\ }\href
  {https://doi.org/10.1103/PhysRevD.92.123006} {\bibfield  {journal} {\bibinfo
  {journal} {Phys. Rev. D}\ }\textbf {\bibinfo {volume} {92}},\ \bibinfo
  {pages} {123006} (\bibinfo {year} {2015})},\ \Eprint
  {https://arxiv.org/abs/1509.00481} {arXiv:1509.00481 [hep-ph]} \BibitemShut
  {NoStop}%
\bibitem [{\citenamefont {Tulin}\ and\ \citenamefont
  {Yu}(2018)}]{Tulin:2017ara}%
  \BibitemOpen
  \bibfield  {author} {\bibinfo {author} {\bibfnamefont {S.}~\bibnamefont
  {Tulin}}\ and\ \bibinfo {author} {\bibfnamefont {H.-B.}\ \bibnamefont {Yu}},\
  }\bibfield  {title} {\bibinfo {title} {{Dark Matter Self-interactions and
  Small Scale Structure}},\ }\href
  {https://doi.org/10.1016/j.physrep.2017.11.004} {\bibfield  {journal}
  {\bibinfo  {journal} {Phys. Rept.}\ }\textbf {\bibinfo {volume} {730}},\
  \bibinfo {pages} {1} (\bibinfo {year} {2018})},\ \Eprint
  {https://arxiv.org/abs/1705.02358} {arXiv:1705.02358 [hep-ph]} \BibitemShut
  {NoStop}%
\bibitem [{\citenamefont {Hu}\ \emph {et~al.}(2000)\citenamefont {Hu},
  \citenamefont {Barkana},\ and\ \citenamefont {Gruzinov}}]{Hu:2000ke}%
  \BibitemOpen
  \bibfield  {author} {\bibinfo {author} {\bibfnamefont {W.}~\bibnamefont
  {Hu}}, \bibinfo {author} {\bibfnamefont {R.}~\bibnamefont {Barkana}},\ and\
  \bibinfo {author} {\bibfnamefont {A.}~\bibnamefont {Gruzinov}},\ }\bibfield
  {title} {\bibinfo {title} {{Cold and fuzzy dark matter}},\ }\href
  {https://doi.org/10.1103/PhysRevLett.85.1158} {\bibfield  {journal} {\bibinfo
   {journal} {Phys. Rev. Lett.}\ }\textbf {\bibinfo {volume} {85}},\ \bibinfo
  {pages} {1158} (\bibinfo {year} {2000})},\ \Eprint
  {https://arxiv.org/abs/astro-ph/0003365} {arXiv:astro-ph/0003365}
  \BibitemShut {NoStop}%
\bibitem [{\citenamefont {Hui}\ \emph {et~al.}(2017)\citenamefont {Hui},
  \citenamefont {Ostriker}, \citenamefont {Tremaine},\ and\ \citenamefont
  {Witten}}]{Hui:2016ltb}%
  \BibitemOpen
  \bibfield  {author} {\bibinfo {author} {\bibfnamefont {L.}~\bibnamefont
  {Hui}}, \bibinfo {author} {\bibfnamefont {J.~P.}\ \bibnamefont {Ostriker}},
  \bibinfo {author} {\bibfnamefont {S.}~\bibnamefont {Tremaine}},\ and\
  \bibinfo {author} {\bibfnamefont {E.}~\bibnamefont {Witten}},\ }\bibfield
  {title} {\bibinfo {title} {{Ultralight scalars as cosmological dark
  matter}},\ }\href {https://doi.org/10.1103/PhysRevD.95.043541} {\bibfield
  {journal} {\bibinfo  {journal} {Phys. Rev. D}\ }\textbf {\bibinfo {volume}
  {95}},\ \bibinfo {pages} {043541} (\bibinfo {year} {2017})},\ \Eprint
  {https://arxiv.org/abs/1610.08297} {arXiv:1610.08297 [astro-ph.CO]}
  \BibitemShut {NoStop}%
\bibitem [{\citenamefont {Marsh}(2016)}]{Marsh:2015xka}%
  \BibitemOpen
  \bibfield  {author} {\bibinfo {author} {\bibfnamefont {D.~J.~E.}\
  \bibnamefont {Marsh}},\ }\bibfield  {title} {\bibinfo {title} {{Axion
  Cosmology}},\ }\href {https://doi.org/10.1016/j.physrep.2016.06.005}
  {\bibfield  {journal} {\bibinfo  {journal} {Phys. Rept.}\ }\textbf {\bibinfo
  {volume} {643}},\ \bibinfo {pages} {1} (\bibinfo {year} {2016})},\ \Eprint
  {https://arxiv.org/abs/1510.07633} {arXiv:1510.07633 [astro-ph.CO]}
  \BibitemShut {NoStop}%
\bibitem [{\citenamefont {Ferreira}(2021)}]{Ferreira:2020fam}%
  \BibitemOpen
  \bibfield  {author} {\bibinfo {author} {\bibfnamefont {E.~G.~M.}\
  \bibnamefont {Ferreira}},\ }\bibfield  {title} {\bibinfo {title}
  {{Ultra-light dark matter}},\ }\href
  {https://doi.org/10.1007/s00159-021-00135-6} {\bibfield  {journal} {\bibinfo
  {journal} {Astron. Astrophys. Rev.}\ }\textbf {\bibinfo {volume} {29}},\
  \bibinfo {pages} {7} (\bibinfo {year} {2021})},\ \Eprint
  {https://arxiv.org/abs/2005.03254} {arXiv:2005.03254 [astro-ph.CO]}
  \BibitemShut {NoStop}%
\bibitem [{\citenamefont {Hui}(2021)}]{Hui:2021tkt}%
  \BibitemOpen
  \bibfield  {author} {\bibinfo {author} {\bibfnamefont {L.}~\bibnamefont
  {Hui}},\ }\bibfield  {title} {\bibinfo {title} {{Wave Dark Matter}},\ }\href
  {https://doi.org/10.1146/annurev-astro-120920-010024} {\bibfield  {journal}
  {\bibinfo  {journal} {Ann. Rev. Astron. Astrophys.}\ }\textbf {\bibinfo
  {volume} {59}},\ \bibinfo {pages} {247} (\bibinfo {year} {2021})},\ \Eprint
  {https://arxiv.org/abs/2101.11735} {arXiv:2101.11735 [astro-ph.CO]}
  \BibitemShut {NoStop}%
\bibitem [{\citenamefont {Roszkowski}\ \emph {et~al.}(2018)\citenamefont
  {Roszkowski}, \citenamefont {Sessolo},\ and\ \citenamefont
  {Trojanowski}}]{Roszkowski:2017nbc}%
  \BibitemOpen
  \bibfield  {author} {\bibinfo {author} {\bibfnamefont {L.}~\bibnamefont
  {Roszkowski}}, \bibinfo {author} {\bibfnamefont {E.~M.}\ \bibnamefont
  {Sessolo}},\ and\ \bibinfo {author} {\bibfnamefont {S.}~\bibnamefont
  {Trojanowski}},\ }\bibfield  {title} {\bibinfo {title} {{WIMP dark matter
  candidates and searches\textemdash{}current status and future prospects}},\
  }\href {https://doi.org/10.1088/1361-6633/aab913} {\bibfield  {journal}
  {\bibinfo  {journal} {Rept. Prog. Phys.}\ }\textbf {\bibinfo {volume} {81}},\
  \bibinfo {pages} {066201} (\bibinfo {year} {2018})},\ \Eprint
  {https://arxiv.org/abs/1707.06277} {arXiv:1707.06277 [hep-ph]} \BibitemShut
  {NoStop}%
\bibitem [{\citenamefont {Khoury}(2022)}]{Khoury:2021tvy}%
  \BibitemOpen
  \bibfield  {author} {\bibinfo {author} {\bibfnamefont {J.}~\bibnamefont
  {Khoury}},\ }\bibfield  {title} {\bibinfo {title} {{Dark Matter
  Superfluidity}},\ }\href {https://doi.org/10.21468/SciPostPhysLectNotes.42}
  {\bibfield  {journal} {\bibinfo  {journal} {SciPost Phys. Lect. Notes}\
  }\textbf {\bibinfo {volume} {42}},\ \bibinfo {pages} {1} (\bibinfo {year}
  {2022})},\ \Eprint {https://arxiv.org/abs/2109.10928} {arXiv:2109.10928
  [astro-ph.CO]} \BibitemShut {NoStop}%
\bibitem [{\citenamefont {Yeh}\ \emph {et~al.}(2022)\citenamefont {Yeh},
  \citenamefont {Shelton}, \citenamefont {Olive},\ and\ \citenamefont
  {Fields}}]{Yeh:2022heq}%
  \BibitemOpen
  \bibfield  {author} {\bibinfo {author} {\bibfnamefont {T.-H.}\ \bibnamefont
  {Yeh}}, \bibinfo {author} {\bibfnamefont {J.}~\bibnamefont {Shelton}},
  \bibinfo {author} {\bibfnamefont {K.~A.}\ \bibnamefont {Olive}},\ and\
  \bibinfo {author} {\bibfnamefont {B.~D.}\ \bibnamefont {Fields}},\ }\bibfield
   {title} {\bibinfo {title} {{Probing physics beyond the standard model:
  limits from BBN and the CMB independently and combined}},\ }\href
  {https://doi.org/10.1088/1475-7516/2022/10/046} {\bibfield  {journal}
  {\bibinfo  {journal} {JCAP}\ }\textbf {\bibinfo {volume} {10}},\ \bibinfo
  {pages} {046}},\ \Eprint {https://arxiv.org/abs/2207.13133} {arXiv:2207.13133
  [astro-ph.CO]} \BibitemShut {NoStop}%
\bibitem [{\citenamefont {Di~Valentino}\ \emph {et~al.}(2021)\citenamefont
  {Di~Valentino}, \citenamefont {Mena}, \citenamefont {Pan}, \citenamefont
  {Visinelli}, \citenamefont {Yang}, \citenamefont {Melchiorri}, \citenamefont
  {Mota}, \citenamefont {Riess},\ and\ \citenamefont
  {Silk}}]{DiValentino:2021izs}%
  \BibitemOpen
  \bibfield  {author} {\bibinfo {author} {\bibfnamefont {E.}~\bibnamefont
  {Di~Valentino}}, \bibinfo {author} {\bibfnamefont {O.}~\bibnamefont {Mena}},
  \bibinfo {author} {\bibfnamefont {S.}~\bibnamefont {Pan}}, \bibinfo {author}
  {\bibfnamefont {L.}~\bibnamefont {Visinelli}}, \bibinfo {author}
  {\bibfnamefont {W.}~\bibnamefont {Yang}}, \bibinfo {author} {\bibfnamefont
  {A.}~\bibnamefont {Melchiorri}}, \bibinfo {author} {\bibfnamefont {D.~F.}\
  \bibnamefont {Mota}}, \bibinfo {author} {\bibfnamefont {A.~G.}\ \bibnamefont
  {Riess}},\ and\ \bibinfo {author} {\bibfnamefont {J.}~\bibnamefont {Silk}},\
  }\bibfield  {title} {\bibinfo {title} {{In the realm of the Hubble
  tension\textemdash{}a review of solutions}},\ }\href
  {https://doi.org/10.1088/1361-6382/ac086d} {\bibfield  {journal} {\bibinfo
  {journal} {Class. Quant. Grav.}\ }\textbf {\bibinfo {volume} {38}},\ \bibinfo
  {pages} {153001} (\bibinfo {year} {2021})},\ \Eprint
  {https://arxiv.org/abs/2103.01183} {arXiv:2103.01183 [astro-ph.CO]}
  \BibitemShut {NoStop}%
\bibitem [{\citenamefont {Hu}\ and\ \citenamefont {Wang}(2023)}]{Hu:2023jqc}%
  \BibitemOpen
  \bibfield  {author} {\bibinfo {author} {\bibfnamefont {J.-P.}\ \bibnamefont
  {Hu}}\ and\ \bibinfo {author} {\bibfnamefont {F.-Y.}\ \bibnamefont {Wang}},\
  }\bibfield  {title} {\bibinfo {title} {{Hubble Tension: The Evidence of New
  Physics}},\ }\href {https://doi.org/10.3390/universe9020094} {\bibfield
  {journal} {\bibinfo  {journal} {Universe}\ }\textbf {\bibinfo {volume} {9}},\
  \bibinfo {pages} {94} (\bibinfo {year} {2023})},\ \Eprint
  {https://arxiv.org/abs/2302.05709} {arXiv:2302.05709 [astro-ph.CO]}
  \BibitemShut {NoStop}%
\bibitem [{\citenamefont {Kamionkowski}\ and\ \citenamefont
  {Riess}(2023)}]{Kamionkowski:2022pkx}%
  \BibitemOpen
  \bibfield  {author} {\bibinfo {author} {\bibfnamefont {M.}~\bibnamefont
  {Kamionkowski}}\ and\ \bibinfo {author} {\bibfnamefont {A.~G.}\ \bibnamefont
  {Riess}},\ }\bibfield  {title} {\bibinfo {title} {{The Hubble Tension and
  Early Dark Energy}},\ }\href
  {https://doi.org/10.1146/annurev-nucl-111422-024107} {\bibfield  {journal}
  {\bibinfo  {journal} {Ann. Rev. Nucl. Part. Sci.}\ }\textbf {\bibinfo
  {volume} {73}},\ \bibinfo {pages} {153} (\bibinfo {year} {2023})},\ \Eprint
  {https://arxiv.org/abs/2211.04492} {arXiv:2211.04492 [astro-ph.CO]}
  \BibitemShut {NoStop}%
\bibitem [{\citenamefont {Alexander}\ \emph {et~al.}(2019)\citenamefont
  {Alexander}, \citenamefont {Bramburger},\ and\ \citenamefont
  {McDonough}}]{Alexander:2019qsh}%
  \BibitemOpen
  \bibfield  {author} {\bibinfo {author} {\bibfnamefont {S.}~\bibnamefont
  {Alexander}}, \bibinfo {author} {\bibfnamefont {J.~J.}\ \bibnamefont
  {Bramburger}},\ and\ \bibinfo {author} {\bibfnamefont {E.}~\bibnamefont
  {McDonough}},\ }\bibfield  {title} {\bibinfo {title} {{Dark Disk Substructure
  and Superfluid Dark Matter}},\ }\href
  {https://doi.org/10.1016/j.physletb.2019.134871} {\bibfield  {journal}
  {\bibinfo  {journal} {Phys. Lett. B}\ }\textbf {\bibinfo {volume} {797}},\
  \bibinfo {pages} {134871} (\bibinfo {year} {2019})},\ \Eprint
  {https://arxiv.org/abs/1901.03694} {arXiv:1901.03694 [astro-ph.CO]}
  \BibitemShut {NoStop}%
\bibitem [{\citenamefont {Alexander}\ \emph {et~al.}(2020)\citenamefont
  {Alexander}, \citenamefont {Gleyzer}, \citenamefont {McDonough},
  \citenamefont {Toomey},\ and\ \citenamefont {Usai}}]{Alexander:2019puy}%
  \BibitemOpen
  \bibfield  {author} {\bibinfo {author} {\bibfnamefont {S.}~\bibnamefont
  {Alexander}}, \bibinfo {author} {\bibfnamefont {S.}~\bibnamefont {Gleyzer}},
  \bibinfo {author} {\bibfnamefont {E.}~\bibnamefont {McDonough}}, \bibinfo
  {author} {\bibfnamefont {M.~W.}\ \bibnamefont {Toomey}},\ and\ \bibinfo
  {author} {\bibfnamefont {E.}~\bibnamefont {Usai}},\ }\bibfield  {title}
  {\bibinfo {title} {{Deep Learning the Morphology of Dark Matter
  Substructure}},\ }\href {https://doi.org/10.3847/1538-4357/ab7925} {\bibfield
   {journal} {\bibinfo  {journal} {Astrophys. J.}\ }\textbf {\bibinfo {volume}
  {893}},\ \bibinfo {pages} {15} (\bibinfo {year} {2020})},\ \Eprint
  {https://arxiv.org/abs/1909.07346} {arXiv:1909.07346 [astro-ph.CO]}
  \BibitemShut {NoStop}%
\bibitem [{\citenamefont {Alexander}\ \emph {et~al.}(2022)\citenamefont
  {Alexander}, \citenamefont {Capanelli}, \citenamefont {G.~M.~Ferreira},\ and\
  \citenamefont {McDonough}}]{Alexander:2021zhx}%
  \BibitemOpen
  \bibfield  {author} {\bibinfo {author} {\bibfnamefont {S.}~\bibnamefont
  {Alexander}}, \bibinfo {author} {\bibfnamefont {C.}~\bibnamefont
  {Capanelli}}, \bibinfo {author} {\bibfnamefont {E.}~\bibnamefont
  {G.~M.~Ferreira}},\ and\ \bibinfo {author} {\bibfnamefont {E.}~\bibnamefont
  {McDonough}},\ }\bibfield  {title} {\bibinfo {title} {{Cosmic filament spin
  from dark matter vortices}},\ }\href
  {https://doi.org/10.1016/j.physletb.2022.137298} {\bibfield  {journal}
  {\bibinfo  {journal} {Phys. Lett. B}\ }\textbf {\bibinfo {volume} {833}},\
  \bibinfo {pages} {137298} (\bibinfo {year} {2022})},\ \Eprint
  {https://arxiv.org/abs/2111.03061} {arXiv:2111.03061 [astro-ph.CO]}
  \BibitemShut {NoStop}%
\end{thebibliography}%

\end{document}